\documentclass[english,aps,pra,amsmath,twocolumn,preprintnumber,superscriptaddress,showpacs,notitlepage]{revtex4-2}
\usepackage[T1]{fontenc}
\usepackage[latin9]{inputenc}
\makeatletter

\usepackage{mathptmx,xspace}
\usepackage{amsbsy,bm}
\usepackage{graphicx,color,xcolor,epsfig,rotate}
\usepackage{fancyhdr}
\usepackage{amssymb}
\usepackage[colorlinks=true, 
            linkcolor=blue, 
            urlcolor=blue,
            citecolor=blue]{hyperref}
\usepackage{soul}
\usepackage{cancel}
\usepackage{amsmath}
\newcommand{\ssz}{\textcolor{blue}}

\makeatother

\usepackage{babel}

\begin{document}

\title{Schwinger boson theory  of ordered magnets}
\author{Shang-Shun~Zhang}
\affiliation{Department of Physics and Astronomy, The University of Tennessee,
Knoxville, Tennessee 37996, USA}
\affiliation{School of Physics and Astronomy and William I. Fine Theoretical Physics
Institute, University of Minnesota, Minneapolis, MN 55455, USA}
\author{E. A. Ghioldi}
\affiliation{Department of Physics and Astronomy, The University of Tennessee, Knoxville, Tennessee 37996, USA}
\author{L. O. Manuel}
\affiliation{Instituto de F{\'i}sica Rosario (CONICET) and Universidad Nacional de Rosario, Boulevard 27 de Febrero 210 bis, (2000) Rosario, Argentina}
\author{A. E. Trumper}
\affiliation{Instituto de F{\'i}sica Rosario (CONICET) and Universidad Nacional de Rosario, Boulevard 27 de Febrero 210 bis, (2000) Rosario, Argentina}
\author{Cristian~D.~Batista}
\affiliation{Department of Physics and Astronomy, The University of Tennessee,
Knoxville, Tennessee 37996, USA}
\affiliation{Quantum Condensed Matter Division and Shull-Wollan Center, Oak Ridge
National Laboratory, Oak Ridge, Tennessee 37831, USA}

\date{\today}
\begin{abstract}

The Schwinger boson theory provides a natural path for treating quantum spin systems with large quantum fluctuations. In contrast to semi-classical treatments, this theory allows us to describe a continuous transition between magnetically ordered and spin liquid states, as well as the continuous evolution of the corresponding excitation spectrum. The square lattice Heisenberg antiferromagnet is one of the first models that was approached with the Schwinger boson theory. Here we revisit this problem to reveal several subtle points that were omitted in previous treatments and that are crucial to  further develop this formalism. These points include the freedom for the choice of the saddle point (Hubbard-Stratonovich decoupling and choice of the condensate) and the $1/N$ expansion in the presence of a condensate. A key observation is that the spinon condensate leads to  Feynman diagrams that include contributions of different order in $1/N$,  which must be accounted  to get a qualitatively correct excitation spectrum. We demonstrate that a proper treatment of these contributions leads to an exact cancellation of the single-spinon poles of the dynamical spin structure factor, as expected for a magnetically ordered state. The only surviving poles are the ones arising from the magnons (two-spinon bound states), which are the true collective modes of an ordered magnet. 

\end{abstract}
\pacs{~}
\maketitle

\section{Introduction}
\label{Intro}

Spin wave theory (SWT) is the traditional approach to describe the low-temperature properties of magnetically ordered states. This semi-classical approach, that is based on a $1/S$ expansion and  becomes exact in the $S \to \infty$  limit, is indeed adequate to describe the low-energy excitation spectrum of  Heisenberg models, whose magnetically ordered ground states have small quantum fluctuations. As it is well-known, quantum fluctuations can be enhanced by different factors, such as low-spin, low-dimensionality and frustration, which guide the search for  quantum spin liquids in real materials. This ongoing search is revealing multiple examples of magnets whose excitation spectrum is not captured by a $1/S$ expansion despite the fact that their ground state is magnetically ordered. It is then necessary to develop alternative approaches that can capture the strong quantum effects revealed by the excitation spectrum of these materials. 

The Schwinger boson theory (SBT) introduced by Arovas and Auerbach\cite{Arovas1988,book2012interacting} provides an alternative path for modelling magnets with strong quantum fluctuations. 
The Schwinger boson (SB) representation of spin operators allows one to reformulate the Heisenberg model as a quartic Hamiltonian of spin-1/2 bosons  subject to the local constraint of $2S$ SBs per site. The quartic Hamiltonian is expressed in terms of  products of SU(2) invariant  bond operators  that explicitly preserve the rotational invariance of the Heisenberg interaction. Magnetic ordering manifests in this formalism via condensation of the SBs~\cite{Hirsch1989,Sarker1989,Chandra1990,book2012interacting}. Historically, the SBT was motivated by multiple reasons: i) the mean field decoupling or saddle point (SP) expansion preserves the SU(2) symmetry, implying that this approach does not violate Mermin-Wagner's theorem~\cite{Mermin66},
 ii) unfrustrated and highly frustrated Heisenberg models can be studied with the same formalism, which is suitable for describing both spin liquid and magnetically ordered states iii)  fractional excitations of spin liquid states, known as \emph{spinons}, become single-particle excitations coupled to an emergent gauge field, and iv) it can be formulated as a large $N$ (number of flavors of the SBs) theory, implying that observables can be expanded in powers of $1/N$. The widespread use of
 the SBT~\cite{auerbach1988spin,Arovas1988,read1989valence,read1990spin,Read1991,sachdev1991large,Mila1991,Sachdev1992,Ceccatto1993,Raykin1993,Chubukov1994,Lefmann1994,Mattsson1995,Chubukov1995,Starykh1995,Trumper1997,Timm1998,Manuel1999,Burkov2002,Misguich2005,Wang2006,Tchernyshyov_2006,Katsura2008,Flint2009,Mezio2011,Mezio2012,Langari2012,Messio2013,Ghioldi2015,yang2016schwinger,Punk2016,Bauer17,ghioldi2018dynamical,zhang2019large,Gang2019} over the last 30 years is a natural consequence of its versatility. However, during the last two decades the interest has mainly focused on quantum spin liquids~\cite{Wen2002,Wen2004,Senthil2004,Misguich2005,Senthil2006,Sachdev2008,Normand2009,Balents2010,Powell2011,Lacroix2011,Yan2011,Han2012,Sorella2012,Hemerle2013,Merino2014,Zhu15,Hu15,Iqbal16,McCulloch2016,Savary2017,Zhou2017,Punk2017,Meng2018,White2018,Takagi19,Wen2019,Becca2019,Gang2019,Broholm2020} and their classification by means of the projective symmetry group theory~\cite{Wen2002,Wen2004,Wang2006,Wang2010,Messio2013,yang2016schwinger,Bauer17,Punk2017,Gang2019}. 
 
 In general, most attempts of using the SBT for describing magnetically ordered and spin liquid states have not gone beyond the saddle-point (SP) level.
 As we will see in this work, this situation could be a natural consequence of several subtleties that have been omitted in previous formulations of the theory and that become particularly important for studying the excitation spectrum of magnetically ordered (or condensed) states. The increasing need of making progress in this front is manifested by a number of magnetically ordered triangular antiferromagnets, such as Ba$_3$CoSb$_2$O$_9$~\cite{Shirata2012,Susuki2013,Ma16,Ito2017,Kamiya,Coldea2020}, whose inelastic neutron scattering cross section cannot be explained
 with a large-$S$ expansion due to their proximity to a ``quantum melting point''~\cite{Ma16,Coldea2020}.

 Motivated by these experimental results, we initiated a zero temperature study of the  $1/N$ corrections of the dynamical spin structure factor of the triangular Heisenberg model. A proper treatment of the condensed phase  -- $120^{\circ}$ N\'eel state-- revealed several technical subtleties that have led  to  misleading interpretations of the SP results and to apparent failures of the SBT~\cite{ghioldi2018dynamical,zhang2019large,zhang2021schwinger}. The most outstanding misinterpretation is the identification of the collective modes with the poles of the dynamical spin structure factor that is obtained at the SP level. We note that these poles coincide with the single-spinon dispersion of the SBT, which in general does not coincide with the single-magnon dispersion. As we have explained in previous works~\cite{ghioldi2018dynamical,zhang2019large}, the inclusion of an additional diagram that would be of order $1/N$ in absence of the condensate leads to an exact cancellation of the SP poles and to the emergence of new poles (zeros of the fluctuation matrix) that represent the true collective modes (magnons) of the system. The correct identification of the true collective modes  allowed us to recover the linear SWT result by taking the large-$S$ limit of the SBT~\cite{zhang2019large}. This identification also explains the failed attempts of recovering the correct large-$S$ limit at the SP level of the SBT~\cite{Chandra1991,Coleman1994}. Unfortunately, this negative result was misinterpreted as an important shortcoming of the SBT that discouraged the use of this formalism for modelling the excitation spectrum of ordered magnets. 
 
 The above-mentioned results have stimulated us to revisit the simpler case of the square Heisenberg model, where the absence of frustration leads to collinear antiferromagnetic ordering at $T=0$ \cite{auerbach1988spin,book2012interacting}. What are our motivations to revisit this old problem? Our first motivation is to understand the differences between different SP approximations that can be used to describe \emph{the same} magnetically ordered state. This freedom is not only related to different ways of decoupling the Heisenberg Hamiltonian into products of two bond operators, but also to the choice of the condensate for a particular decoupling scheme. As we will see in this work, the differences between different decoupling schemes parametrized by the continuous parameter $\alpha$ become less important upon including higher order corrections in $1/N$. In particular, we will see that 
 the dynamical spin structure factor coincides with the linear spin wave result in the $S \rightarrow \infty$ limit {\it for any value of $\alpha$}.
 As for the choice of the condensate, we will see that different condensates that describe the same long-range magnetic ordering lead to different low-order $1/N$ corrections of the dynamical spin susceptibility. 
 Correspondingly, at the SP level, there is an optimal choice of the condensate that can be continuously connected with the \emph{simple} Bose-Einstein condensate (BEC) solution (all bosons are condensed in the same mode) \cite{leggett2006quantum} obtained for non-collinear orderings (the square lattice can be continuously deformed into a triangular lattice).

 

The second important motivation is to revisit the diagramatic $1/N$ expansion \cite{auerbach1988spin,book2012interacting} in the presence of a condensate. In particular, we will demonstrate that each Feynman diagram of the dynamical spin susceptibility $\chi(\bf k, \omega)$ has contributions of different order in $1/N$ in the presence of a condensate. More importantly, this mixed character of each Feynman diagram leads to an exact cancellation of the single-spinon poles to any order in $1/N$. In other words, for each diagram that includes single spinon poles, there is a counter-diagram that removes these spurious poles from the excitation spectrum of a magnetically ordered state. Since the \emph{non-condensed} parts of the diagram and the counter-diagram are of different order in $1/N$, this result  forced us to reconsider the diagramatic hierarchy in the presence of a condensate: for each diagram that is included in the calculation, one must also include the corresponding counter diagram to obtain physically correct results. In particular, this explains why the SP result is in general not enough to obtain a qualitatively correct dynamical spin structure factor.
From a technical point of view, we uncover the unusual  structure of the $1/N$ diagrammatic expansion of the dynamical spin susceptibility in the condensed phase. Namely, diagrams of $\chi_{FL}(\bf k, \omega)$ 
that are nominally of order $1/N$ and account for the fluctuations around the SP solution include a singular contribution of order $O(1/N^0)$ whenever the condensate fraction is finite. This contribution corresponds to isolated poles with residues of order $O(1/N^0)$ that exactly cancel the single spinon poles of the saddle point solution $\chi_{SP}(\bf k, \omega)$. In general, for each diagram we identify the counter-diagram  that must be added in to cancel the unphysical single-spinon poles. 
 
 The article is organized as follows. In Sec.~\ref{SBR} we introduce the SB representation of the Heisenberg model along with its generalization to arbitrary number $N$ of bosonic flavors. We also introduce the most general decoupling scheme of the spin Hamiltonian in terms of products of bond operators.
 Sec.~\ref{PIF} is devoted to the path integral formulation based on the Hubbard-Stratonvich transformation associated with the bond operator decoupling described in Sec.~\ref{SBR}.  The SP approximation and the corresponding SP solution (spinon condensation associated with long-range magnetic ordering) are discussed in Sec.~\ref{SP}.
 This section also includes a discussion of the different ways in which the SBs can be condensed in the thermodynamic limit. 
 In Sec.~\ref{BSP} we introduce the formalism required to go beyond the SP level and in Sec.~\ref{CSSP} we demonstrate the non-trivial cancellation of the single-spinon poles of the dynamical spin structure factor. In particular, Sec.~\ref{CSSP} includes a careful derivation of the diagrammatic expansion in the presence of a condensate. The emergence of the true collective modes (magnons) of the theory is presented in Sec.~\ref{MP}. This section also includes a comparison between the results that are obtained for different values of $\alpha$. The main results of the manuscript are summarized in Sec.~\ref{SUM}.

\section{Schwinger boson representation}
\label{SBR}

We will consider the antiferromagnetic (AFM) Heisenberg model 
\begin{equation}
{\hat H}_{AFM}=\sum_{\langle i,j \rangle} J_{ij} {\hat{\bm{S}}}_{i}\cdot {\hat{\bm{S}}}_{j}, \label{eq:Hxxz}
\end{equation}
where $\hat{\bm{S}_{i}}$ is the spin operator at the lattice site $i$ and $J_{ij}>0$ is the antiferromagnetic exchange constant.
For a bipartite lattice~\cite{auerbach1988spin,read1989valence,read1990spin}, this Hamiltonian can be generalized 
by replacing the spin operators, which are generators of the SU(2) group, with the generators of the SU(N) group. 
For the generalization of the AFM Heisenberg model, we must use  conjugate representations of  SU(N) for each of the two $A$ and $B$ sublattices~\footnote{The conjugate representations are required to have a singlet ground state of the two-site Heisenberg Hamiltonian.}.
In terms of an $N$-component Schwinger boson,
${\hat{b}}_i=({\hat{b}}_{i,1},...,{\hat{b}}_{i,N})^T$, the generators of the  SU(N) group can be expressed as ${\hat{S}}^{mn}_i = {\hat{b}}_{i,m}^{\dagger} {\hat{b}}_{i,n}^{\;}$
on the  $A$ sublattice and  ${\hat{S}}^{mn}_i = {\hat{b}}_{i,n}^{\dagger} {\hat{b}}_{i,m}^{\;}$ on the  $B$ sublattice~\cite{auerbach1988spin,book2012interacting,read1989valence}.
The constraint equation:
\begin{equation}
\sum_{m} {\hat{b}}_{i,m}^{\dagger}{\hat{b}}_{i,m} = N S,
\label{constraint}
\end{equation}
where $N S$ is an integer number, determines a particular  representation of SU(N).
In other words,  $S$ represents the spin size for $N=2$.

Because of the requirement of conjugate representations for the two different sublattices, this SU(N) extension does not apply to Heisenberg antiferromagnets on non-bipartite lattices. A more ``flexible'' generalization of the spin operator is provided by the generators of the $Sp(N/2)$ group, for even values of $N$, because  all spin operators belong to the same representation~\cite{sachdev1991large}.
The corresponding generalization of the AFM Heisenberg Hamiltonian is 
\begin{equation}\label{eq:spn}
{\hat H}_{Sp(N/2)} = \sum_{\langle i,j \rangle} \frac{4 J_{ij}}{N} \left( \frac{N S^2}{4} - {\hat A}_{ij}^{\dagger}{\hat A}_{ij} \right).
\end{equation}
The operator
\begin{eqnarray}\label{bond_op2}
{\hat A}_{ij}^{ \dag} = \frac{1}{2}\sum_{\sigma = -N/2}^{N/2} \text{sign}(\sigma) {\hat{b}}_{ i {\bar  \sigma}}^{ \dag} {\hat{b}}_{ j \sigma}^{ \dag},
\label{Aij}
\end{eqnarray}
with ${\bar \sigma}=- \sigma$, creates a singlet state that is invariant under the group of $Sp(N/2)$ transformations. 

There is still another  generalization of the AFM Heisenberg model based on the so-called ``symplectic spin''~\cite{Flint2009}, which preserves the odd nature of the spin operator under time-reversal transformation. The generalized spin operators 
generate a subgroup of SU(N), dubbed as $Sy(N/2)$ in this work. The corresponding  Hamiltonian is
\begin{equation}\label{eq:syn}
{\hat H}_{Sy(N/2)} = \sum_{\langle i,j \rangle} \frac{2 J_{ij}}{N} \left( :{\hat B}_{ij}^{\dagger}{\hat B}_{ij}: - {\hat A}_{ij}^{\dagger}{\hat A}_{ij} \right),
\end{equation}
where the bond operator $\hat{A}_{ij}^{\dag}$ has been defined in Eq.~\eqref{Aij} and
\begin{eqnarray}\label{bond_op3}
{\hat B}_{ij}^{ \dag} = \frac{1}{2} \sum_{\sigma= -N/2}^{N/2} {\hat{b}}_{ i \sigma }^{\dagger} {\hat{b}}_{ j \sigma}^{}.
\end{eqnarray}

Although these generalizations of the AFM Heisenberg model are different for  $N>2$, they coincide for the physical case of interest ($N=2$) because $Sp(1)$ and $Sy(1)$ are isomorphic to SU(2). The following identity holds for this particular ($N=2$) case:
\begin{eqnarray}\label{identity}
:{\hat B}_{ij}^{\dagger}{\hat B}_{ij}:  + {\hat A}_{ij}^{\dagger}{\hat A}_{ij} = S^2.
\label{apb}
\end{eqnarray}
This identity allows us to express the $N=2$ AFM Heisenberg Hamiltonian in multiple ways that are parametrized by the continuous parameter  $\alpha$: 
\begin{eqnarray}
\! \! \! \! {\hat H}_{SB} (\alpha,N) \!  \! = \! \! \sum_{\langle ij\rangle} \! \frac{4 J_{ij}}{N} \!\! \left[ C_\alpha + \alpha :{\hat B}_{ij}^{\dagger}{\hat B}_{ij}:- (1-\alpha) {\hat A}_{ij}^{\dagger}{\hat A}_{ij}\right]\!\! \!, \label{eq:SBlargeN}
\end{eqnarray}
where $C_\alpha=N  S^2 (1/4-\alpha/2)$.
In other words,   ${\hat H}_{SB} (\alpha,N=2)$ is  independent of $\alpha$  up to an irrelevant additive constant.
As we will see in this work, while the different mean field decouplings of ${\hat H}_{SB}$ associated with each value of the continuous parameter  $\alpha$ lead to very different SP solutions, the  inclusion of a $1/N$ correction makes the dynamical spin structure factor (DSSF) much less dependent on $\alpha$. 
In particular, we will demonstrate that  the  DSSF coincides with the linear spin wave result in the large $S \to \infty$ limit \emph{for any} value of $\alpha$.

For concreteness, we will focus on the  square lattice. As we will see later, the collinear nature of the AFM ordering in this lattice introduces an extra freedom in the choice of the spinon condensate. The rest of the analysis, that is included in the next sections, can be extended to other lattice/models, including non-bipartite lattices with non-collinear magnetic orderings.

\section{Path integral formulation}
\label{PIF}

Eq.~\eqref{eq:SBlargeN} provides a natural decoupling scheme to perform the Hubbard-Stratonovich transformation
that is the basis of the SBT. As we already mentioned, due to the singlet character of the bond operators ${\hat A}_{ij}$ and ${\hat B}_{ij}$, the SP equations, or equivalently the mean field Hamiltonian, produced by this decoupling scheme do not explicitly break the SU(2) symmetry of the Heisenberg Hamiltonian. The basic steps for implementing the path integral over coherent states can be found in Refs.~~\cite{Negele1998,book2012interacting}.

After performing the Hubbard-Stratonovich transformation, the partition function of the model Hamiltonian Eq.~(\ref{eq:SBlargeN}) can be expressed as
 the path-integral~\cite{ghioldi2018dynamical}
\begin{equation}
{\cal Z}=\int{\cal D}[\bar{W}W\lambda]\int{\cal D}[\bar{b}b]\exp(-N S(\bar{W},W,\lambda,\bar{b},b)).
\end{equation}
$\bar{W},W$ are the Hubbard-Stratonovich (HS) complex fields,   $b$ is the complex field of eigenvalues of the annihilation SB operators $\hat{b} $ (the corresponding eigenkets are coherent states), and  $\lambda$ is a real field introduced to enforce the constraint equation \eqref{constraint}. The action $S(\bar{W},W,\lambda,\bar{b},b)$ can be decomposed into three terms:
\begin{align} \label{action0}
S(\bar{W},W,\lambda,\bar{b},b) = &S_{0}(\bar{W},W,\lambda)+S_{b\lambda}(\bar{b},b, \lambda) \nonumber \\ &
+S_{b W}(\bar{b},b,\bar{W},W,),
\end{align}
with
\begin{multline}
S_{0}(\bar{W},W,\lambda) = \int_{0}^{\beta}d\tau\sum_{\langle ij\rangle}\frac{1}{4J_{ij}} [ \alpha \bar{W}_{ij}^{B}(\tau)W_{ij}^{B}(\tau) \\
+ (1-\alpha) \bar{W}_{ij}^{A}(\tau)W_{ij}^{A}(\tau) ] - i S\int_{0}^{\beta}d\tau\sum_{i}\lambda_{i}(\tau),
\end{multline}
\begin{eqnarray}
S_{b \lambda}(\bar{b},b, \lambda) &=&
\frac{1}{2N}\int_{0}^{\beta}d\tau\sum_{i} \psi_{i}^{\dagger}(\partial_{\tau}\gamma^0+i\lambda_{i}(\tau))\psi_{i},
\end{eqnarray}
\begin{multline}
S_{b W}(\bar{b},b,\bar{W},W) =  \frac{\alpha}{N} \int_{0}^{\beta}d\tau\sum_{\langle ij\rangle} 
[ {\psi}^{\dagger}_{i}(\Gamma_{B}^{} \bar W_{ji}^{B} - \Gamma_{B}^{\dagger}W_{ij}^{B})\psi_{j} ] \\
-\frac{1- \alpha }{N} \int_{0}^{\beta}d\tau\sum_{\langle ij\rangle}  [ {\psi}^{\dagger}_{i} (\Gamma_{A}\bar{W}_{ji}^{A}+\Gamma_{A}^\dag {W}_{ij}^{A})\psi_{j}]. 
\end{multline}
where $\beta$ is the inverse temperature, ${\psi}^{\dagger}_{i}=(\bar{b}_{i,N/2},...,\bar{b}_{i,-N/2},b_{i,N/2},...,b_{i,-N/2})$ is the $2N$-component Nambu-spinor, the superindex $\mu$ of $\bar{W}_{ij}^{\mu}$ and $W_{ij}^{\mu}$denotes the auxiliary field that decouples the $\bar{A}_{ij}A_{ij}$ and $\bar {B}_{ij} {B}_{ij}$ terms of the Hamiltonian Eq.~\eqref{eq:SBlargeN}, and
$\lambda_i$ is a Langrange multiplier that enforces the local  constraint Eq.~\eqref{constraint} on each site.
The constant matrix $\gamma^0$ is defined by $\gamma^0\equiv \sigma_z \otimes I_0$, where $I_0$ is the $N\times N$ identity matrix.
The bond fields can be expressed in terms of the Nambu-spinor as 
\begin{equation}
{A}_{ij}=\psi_{j}^{\dagger}\Gamma_{A}\psi_{i},
\;\;\;
{ B}_{ij}=\psi_{j}^{\dagger}\Gamma_{B}\psi_{i},
\end{equation}
where  $\Gamma_{A}$ and $\Gamma_{B}$ are the matrices of the bilinear forms  given in Eqs.~\eqref{bond_op2} and \eqref{bond_op3}, respectively.

The Nambu representation has an artificial ``particle-hole'' symmetry because $\psi_{i}=  P  ( \psi_{i}^{\dagger}  )^{T}$,
where $P=\sigma_{x}\otimes I_{0}$.  
We can then rewrite the bond fields in the more symmetric form
\begin{align}
{B}_{ij} & =\frac{1}{2}\psi_{j}^{\dagger}\Gamma_{B}\psi_{i}+\frac{1}{2}\psi_{i}^{\dagger}\Upsilon_{B}\psi_{j},\\
{A}_{ij} & =\frac{1}{2}\psi_{j}^{\dagger}\Gamma_{A}\psi_{i}+\frac{1}{2}\psi_{i}^{\dagger}\Upsilon_{A}\psi_{j},
\end{align}
where $\Upsilon_{B}=P\Gamma_{B}^{T}P$ and $\Upsilon_{A}=P\Gamma_{A}^{T}P$.
The effective action \eqref{action0} is invariant under the $U(1)$ gauge transformation
$b_{i\sigma}(\tau) \rightarrow b_{i\sigma}(\tau) e^{i \theta_{i}(\tau)}$ if the auxiliary fields transform as
\begin{eqnarray}
 W_{ij}^{\mu}(\tau) &\rightarrow& W_{ij}^{\mu}(\tau) e^{i(\theta_{i}(\tau) \pm \theta_{j}(\tau) )} \nonumber \\
 \overline{W}_{ij}^{\mu}(\tau) &\rightarrow& \overline{W}_{ij}^{\mu}(\tau) e^{-i(\theta_{i}(\tau) \pm \theta_{j}(\tau) )} \label{transf} \\
 \lambda_{ i}(\tau) &\rightarrow& \lambda_{{i}}(\tau) - \partial_{\tau}\theta_{i}(\tau) \nonumber,
\end{eqnarray} 
where the $+$ and $-$ signs hold for $\mu=A$ and $B$, respectively.



The momentum space representation of the action is obtained by Fourier transforming the fields
\begin{align}
\psi_{{i}}^{\dag}(\tau) & =\frac{1}{\sqrt{{\cal N}_{s}\beta}}\sum_{\bm{k},i\omega_{n}} \psi^{\dag}(\bm{k},i\omega_{n})C_{1}(i\omega_{n})e^{-i\left(\bm{k}\cdot\bm{r}_i-\omega_n\tau\right)},\\
\psi_{i}(\tau) & =\frac{1}{\sqrt{{\cal N}_{s}\beta}}\sum_{\bm{k},i\omega_{n}}C_{2}(i\omega_{n})\psi(\bm{k},i\omega_{n})e^{i\left(\bm{k}\cdot\bm{r}_i-\omega_n\tau\right)},
\end{align}
\begin{align}
W_{i,i+\bm{\delta}}^{\mu}(\tau) & =\frac{1}{\sqrt{{\cal N}_{s}\beta}}\sum_{\bm{k},i\omega_{n}}e^{i(\bm{k}\cdot\bm{r}_i-\omega_{n}\tau)}W_{\bm{\delta}}^{\mu}(\bm{k},i\omega_{n}),\\
\bar{W}_{i,i+\bm{\delta}}^{\mu}(\tau) & =\frac{1}{\sqrt{{\cal N}_{s}\beta}}\sum_{\bm{k},i\omega_{n}}e^{-i(\bm{k}\cdot\bm{r}_i-\omega_{n}\tau)}\bar{W}_{\bm{\delta}}^{\mu}(\bm{k},i\omega_{n}),\\
\lambda_{i}(\tau) & =\frac{1}{\sqrt{{\cal N}_{s}\beta}}\sum_{\bm{k},i\omega_{n}}e^{i(\bm{k}\cdot\bm{r}_i-\omega_{n}\tau)}\lambda(\bm{k},i\omega_{n}),
\end{align}
where ${\cal N}_s$ is the number of lattice sites. The lattice sites have been labeled by the position vector ${\bm r}_i$ and ${\bm \delta}$ is  the relative vector connecting neighboring sites. The matrices 
\begin{eqnarray*}
C_{1}(i\omega_{n})=\left(\begin{array}{cc}
I_0 e^{i\omega_{n}\eta^{+}} & 0\\
0 & I_0
\end{array}\right), \;\;\; C_{2}(i\omega_{n})=\left(\begin{array}{cc}
I_0 & 0\\
0 & I_0 e^{-i\omega_{n}\eta^{+}}
\end{array}\right).
\end{eqnarray*}
include the convergence factors required to ensure the normal ordering of the spinor fields ${\bar b}, b$ in the action.
The resulting expression of the action is
\begin{widetext}
\begin{equation}
S(\bar{W},W,\lambda,\bar{b},b) =-i S\sqrt{{\cal N}_{s}\beta}\lambda(\bm{0},i0)
+\sum_{\bm{\delta}>0} \frac{1}{4J_{{\bm \delta}}} \sum_{ k } [\alpha \bar{W}_{\bm{\delta}}^{B}(k)W_{\bm{\delta}}^{B}(k) 
+ (1-\alpha) \bar{W}_{\bm{\delta}}^{A}(k)W_{\bm{\delta}}^{A}(k) ] + \frac{1}{2N}\sum_{k,p} {\psi}^{ \dagger}(k){\cal M}(k; p)\psi(p),
\label{ms-action}
\end{equation}
where $k\equiv(\bm{k},i\omega_{n})$, $p\equiv (\bm{p},i\nu_{m})$, the sum $\bm{\delta}>0$ runs over all translation-non-equivalent bonds, and the dynamical matrix
\begin{align}
{\cal M}(k; p) &=  -i\omega_{n}C_{1}(i\omega_{n}) \gamma^0 C_{2}(i\omega_{n}) \delta_{k,p} + \frac{1}{\sqrt{{\cal N}_{s}\beta}} v_{\lambda}(k;p)\lambda( k-p)
\nonumber \\
&+ \frac{1}{\sqrt{{\cal N}_{s}\beta}} \sum_{\bm{\delta}>0} [ 
  v_{W_{\bm{\delta}}^{B}}(k; p) W_{\bm{\delta}}^{B}(k-p)
+ v_{W_{\bm{\delta}}^{A}}(k;p)W_{\bm{\delta}}^{A}(k-p) ] \nonumber \\
&+ \frac{1}{\sqrt{{\cal N}_{s}\beta}}\sum_{\bm{\delta}>0} [ 
 v_{\bar{W}_{\bm{\delta}}^{B}}(k;p) \bar{W}_{{\bm \delta}}^{B}(p-k)
+ v_{\bar{W}_{\bm{\delta}}^{A}}(k;p)\bar{W}_{\bm{\delta}}^{A}(p-k) ] ,
\end{align}
has been expressed in terms of the internal vertices
\begin{eqnarray}
v_{W_{\bm{\delta}}^{A}}( k;p ) & = & \left(\frac{1}{ \sqrt{{\cal N}_{s}\beta}}\right)^{-1} \frac{\delta{\cal M}(k; p)}{\delta W_{\bm{\delta}}^{A}( k-p )} 
= -(1-\alpha)  C_{1}(i\omega_{n})\left(\Gamma_{A}^{\dagger}e^{i\bm{p}\cdot\bm{\delta}}+\Upsilon_{A}^{\dagger}e^{-i\bm{k}\cdot\bm{\delta}}\right)C_{2}(i\nu_{m}), 
\nonumber \\
v_{\bar{W}_{\bm{\delta}}^{A}}( k;p ) & = & \left(\frac{1}{\sqrt{{\cal N}_{s}\beta}}\right)^{-1}\frac{\delta{\cal M}( k; p )}{\delta\bar{W}_{\bm{\delta}}^{A}(p-k)}
=- (1-\alpha) C_{1}(i\omega_{n})  \left(\Gamma_{A}e^{-i\bm{k}\cdot\bm{\delta}}+\Upsilon_{A}e^{i\bm{p}\cdot\bm{\delta}}\right)C_{2}(i\nu_{m}),
\nonumber \\
v_{W_{\bm{\delta}}^{B}}( k;p ) & = & \left(\frac{1}{\sqrt{{\cal N}_{s}\beta}}\right)^{-1}\frac{\delta{\cal M}( k; p )}{\delta W_{\bm{\delta}}^{B}( k-p )}
=- {\alpha} C_{1}(i\omega_{n}) \left(\Gamma_{B}^{\dagger}e^{i\bm{p}\cdot\bm{\delta}}+\Upsilon_{B}^{\dagger}e^{-i\bm{k}\cdot\bm{\delta}}\right)C_{2}(i\nu_{m}),
\nonumber  \\
v_{\bar{W}_{\bm{\delta}}^{B}}( k;p ) & = & \left(\frac{1}{\sqrt{{\cal N}_{s}\beta}}\right)^{-1}\frac{\delta{\cal M}( k; p )}{\delta\bar{W}_{\bm{\delta}}^{B}( p-k )}
= {\alpha} C_{1}(i\omega_{n})  \left(\Gamma_{B}e^{-i\bm{k}\cdot\bm{\delta}}+\Upsilon_{B}e^{i\bm{p}\cdot\bm{\delta}}\right)C_{2}(i\nu_{m}),
\nonumber \\
v_{\lambda}( k;p ) & = & \left(\frac{1}{\sqrt{{\cal N}_{s}\beta}}\right)^{-1}\frac{\delta{\cal M}( k; p )}{\delta\lambda( k-p )}
= iC_{1}(i\omega_{n})C_{2}(i\nu_{m}).
\label{intvert}
\end{eqnarray}
\end{widetext}

The bosonic field can  be integrated out to obtain an effective action in terms of the auxiliary fields 
$\bar{W}$, $W$ and $\lambda$:
\begin{eqnarray} \label{partition}
{\cal Z}= \int{\cal D}[\bar{W}W\lambda]   \exp(-N S_{\rm eff}(\bar{W},W,\lambda)),
\end{eqnarray}
where
\begin{eqnarray}
S_{\rm eff}(\bar{W},W,\lambda) & = & S_{0}[\bar{W},W,\lambda]+\frac{1}{2N}\text{Tr}\ln ({\cal M}).
\label{action2}
\end{eqnarray}
Note that the dynamical matrix ${\cal M}$ acts on the
space $(\xi,\bm{q},\omega)$ where $\xi$ refers to the index of the Nambu spinor.

Since it is not possible to obtain the exact partition function associated with the action~\eqref{action2}, we need to 
introduce an approximation scheme. As usual in these cases, we expand the action in powers of $1/N$ \cite{zhang2019large}.  The lowest order ${\cal O}(1/N^0)$ term corresponds to the SP solution, that becomes exact in the  $N\rightarrow\infty$ limit. 
The SP solution for the effective action, $S_{\rm eff}^{\rm SP}$, is determined from the saddle point conditions $\delta S_{\rm eff}^{\rm SP}/\delta W_{{\bm \delta}}^{\mu}(q)=0$ and 
$\delta S_{\rm eff}^{\rm SP}/\delta \lambda(q)=0$, which lead to
\begin{widetext}
\begin{eqnarray}\label{self-sp-eq1}
\alpha \bar{W}_{\bm{\delta}}^{B}(q)|_{\rm SP}  +  \frac{2 J_{{\bm \delta}}}{\sqrt{{\cal N}_{s}\beta}}\sum_{p} \frac{1}{N} \text{Tr} [{\cal G}_{\rm SP}(p;p+q) v_{W_{\bm{\delta}}^{B}} (p+q;p)] = 0,\\
\label{self-sp-eq2}
(1-\alpha) \bar{W}_{\bm{\delta}}^{A}(q)|_{\rm SP}  +  \frac{2 J_{{\bm \delta}}}{\sqrt{{\cal N}_{s}\beta}}\sum_{p} \frac{1}{N} \text{Tr} [{\cal G}_{\rm SP}(p;p+q) v_{W_{\bm{\delta}}^{A}} (p+q;p)] = 0,\\
\label{self-sp-eq3}
iS \delta_{q,0} - \frac{1}{2{\cal N}_{s}\beta}\sum_{p} \frac{1}{N} \text{Tr}\left[{\cal G}_{\rm SP}(p;p+q) v_{\lambda}(p+q; p) \right] = 0,
\end{eqnarray}
where $q\equiv (\bm{q},i\omega_{n}),p\equiv(\bm{p},i\nu_{m})$, and
\begin{eqnarray}
{\cal G}_{\rm SP}(p; p +q)  =  \frac{1}{{\cal Z}_{ \rm SP}}\int{\cal D}[\bar{b}b]e^{-N S_{\rm eff}(\bar{W}_{\rm SP},W_{\rm SP},\lambda_{ \rm SP},\bar{b},b)}\psi(p){\psi}^{\dagger}(p+q),
\label{eq:green}
\end{eqnarray}
\end{widetext}
is the Schwinger boson's Green's function and ${\cal Z}_{ \rm SP} =\int{\cal D}[\bar{b}b]e^{-N S_{\rm eff}(\bar{W}_{ \rm SP},W_{\rm SP},\lambda_{\rm SP},\bar{b},b)}$. 

\section{Saddle point approximation and spinon condensation}
\label{SP}

The next step is to solve the  self-consistent SP equations [Eqs.~\eqref{self-sp-eq1} to \eqref{self-sp-eq3}]. As we will see in this section, the magnetic ordering of the ground state manifests via the condensation of the SBs in a single particle ground state (zero energy single-spinon mode).  The existence of multiple zero energy single-spinon modes introduces some freedom in the selection of the single-spinon BEC. Part of this freedom is associated with the possible orientations of the uniform magnetization in the twisted reference frame (order parameter)  and it is removed by  the inclusion of an infinitesimal uniform magnetic field $h$ that is sent to zero at the end of the calculation. However, for collinear magnetic orderings there is still a remaining freedom associated with the existence of two zero modes with a given spin polarization. As we will see later, all of these condensates describe the same type of 
 collinear magnetic 
ordering and the choice of ``simple BEC'' (condensation in a unique single-spinon mode) becomes advantageous upon adding corrections beyond the SP approximation. 

For concreteness, we will consider the square lattice  antiferromagnet ($N=2$) with $C_4$ symmetry ( $J_{\bm{\delta}}\equiv J$ ) and the general decomposition 
of the spin Hamiltonian given in Eq.~(\ref{eq:SBlargeN}). Since the ground state of the square lattice AFM Heisenberg model exhibits  N\'{e}el antiferromagnetic order, it is convenient to work in a twisted reference frame, ${{\hat{\bm S}}^{\prime}}_{i\in A} = {\hat {\bm S}}_{i\in A}$ and ${\hat {\bm S}^{\prime}_{i\in B}} = (-{\hat S}_{i}^{x}, {\hat S}_{i}^{y}, -{\hat S}_{i}^{z} )$, where the magnetic order becomes ferromagnetic. Going back to the canonical formalism, the Schwinger boson spinors in the original and new reference frames are
\begin{equation}
{\hat {\bm b}}_{j }  =
\left(\begin{array}{c}
\hat {b}_{j,\uparrow}\\
\hat {b}_{j,\downarrow}
\end{array}\right), \;\;\;
{\hat {\bm b}}^{\prime}_{j }  =
\left(\begin{array}{c}
\hat {b}^{\prime}_{j,\uparrow}\\
\hat {b}^{\prime}_{j,\downarrow}
\end{array}\right),
\end{equation}
respectively. These two spinors are related by the transformation
\begin{equation}\label{rotation}
\hat {{\bm b}}^{\prime}_{j} = \hat {\bm b}_{j } \; {\rm for} \; j\in A, \;\;
\hat {{\bm b}}^{\prime}_{j } = e^{i \frac{\pi}{2} \sigma_y}  \hat {\bm b}_{j} \; {\rm for} \; j\in B,
\end{equation}
which leads to the following transformation of the  bond operators 
\begin{equation}
{\hat A}_{ij} = e^{i\bm{\pi}\cdot\bm{r}_i} {{\hat A}}^{\prime}_{ij}, 
\;\;\;
{\hat B}_{ij} = e^{i\bm{\pi}\cdot\bm{r}_i}{{\hat B}}^{\prime}_{ij},
\end{equation} 
where $\bm{\pi}=(\pi,\pi)$ and 
\begin{equation}
{\hat A}_{ij}^{\prime \dag}= -(1/2){{\hat {\bm b}}^{\prime \dagger}}_{i} ({{{\hat {\bm b}}^{\prime \dagger}}_{j}})^T,
\;\;\;
{\hat {{B}}}^{\prime}_{ij} = (1/2) {\hat {\bm b}}_{j}^{\prime \dagger} (i\sigma^y) {\hat {\bm b}}^{\prime}_{i }.
\end{equation}
The generalization of the transformation in Eq.~\eqref{rotation} to arbitrary values of $N$  is:
\begin{equation}
\hat {{b}}^{\prime}_{j\sigma} = \hat { b}_{j \sigma} \; {\rm for} \; j\in A, \;\; \;\;
\hat {{b}}^{\prime}_{j \sigma } = \text{sign}(\sigma) \hat {b}_{j  {\bar \sigma}} \; {\rm for} \; j\in B.
\end{equation}
From now on, we will work in the twisted frame and the \textit{prime} will be  dropped  to simplify the notation.

Knowing that the ground state ordering  corresponds to a uniform ferromagnetic solution in the twisted reference frame, 
the SP solution of interest must be \emph{uniform and static}. From the projective symmetry group analysis~\cite{yang2016schwinger}, 
the N\'eel antiferromagnetic order can be obtained via spinon  condensation in the $0$-flux phase defined by the SP solution  
\begin{align}
W_{\bm{\delta}}^{B}|_{SP} & =-(\bar{W}_{\bm{\delta}}^{B}|_{SP})^{*}=- {2 J} {\cal B}_{\bm{\delta}},\label{ansatz1-1}\\
W_{\bm{\delta}}^{A}|_{SP} & =(\bar{W}_{\bm{\delta}}^{A}|_{SP})^{*}= {2 J} {\cal A}_{\bm{\delta}},\label{ansatz2-1}\\
\lambda_{\bm{r}}(\tau)|_{SP} & =-i\lambda,
\end{align}
with
\begin{align}
{\cal A}_{(1,0)} & ={\cal A}_{(0,1)}=A_{x},\\
{\cal B}_{(1,0)} & =-{\cal B}_{(0,1)}=iB_{x},
\end{align}
where $A_{x}, B_x$, and $\lambda$ are real numbers. The uniform and static character of the SP solution makes the dynamical matrix diagonal in  momentum and frequency space, ${\cal M}(k; q) = {\cal M}(q) \delta_{k,q}$, where
\begin{align}
{{\cal M}}(q) = {1\over 2}C_{1}(i\omega_{n})  [-i\omega_{n} \gamma^0  + { H}_{\rm sp}({\bm q})] C_{2}(i\omega_{n}), 
\end{align}
and
\begin{eqnarray}
H_{\rm sp}(\bm{q}) & = & \left(\begin{array}{cccc}
\lambda & -i \xi_{\bm{q}} & \Delta_{\bm{q}} & 0\\
i\xi_{\bm{q}} & \lambda & 0 & \Delta_{\bm{q}}\\
\Delta_{\bm{q}} & 0 & \lambda & i \xi_{\bm{q}}\\
0 & \Delta_{\bm{q}} & -i \xi_{\bm{q}} & \lambda,
\end{array}\right)\label{eq:Hmf}
\end{eqnarray}
is the SP Hamiltonian matrix. The operator
\begin{equation}
\hat{{H}}_{SP} = \sum_{\bm q} {\hat \psi}^{\dagger}({\bm q}) H_{\rm sp}({{\bm q}}) {\hat \psi}({{\bm q}})    
\label{Hspq}
\end{equation}
is the corresponding Hamiltonian  in momentum space, where 
\begin{equation}
{\hat \psi}({{\bm q}}) = \frac{1}{\sqrt{{\cal N}_s}}\sum_j {\hat \psi}_j e^{i{\bm q}\cdot{\bm r}_j}
\end{equation}
is the Fourier transform of the Nambu-spinor \emph{operator}
\begin{eqnarray}
{\hat \psi}_j = \left(
\begin{array}{c}
{\hat b}_{j \uparrow} \\
{\hat b}_{ j \downarrow} \\
{\hat b}^{\dagger}_{j \uparrow} \\
{\hat b}^{\dagger}_{j \downarrow} \\
\end{array}
\right ).
\label{fns}
\end{eqnarray}
The matrix elements of $H_{\rm sp}({{\bm q}})$ are defined by
\begin{align}
\Delta_{\bm{q}}=2(1-\alpha)JA_{x}\gamma^{+}_{\bm{q}}, \;\;\;\;\;\; \xi_{\bm{q}}=8\alpha J B_{x}\gamma_{\bm{q}}^{-},
\end{align}
and $\gamma_{\bm{q}}^{\pm} =\frac{1}{2}\left(\cos{q_{x}} \pm \cos{q_{y}} \right)$.

\subsection{Symmetry analysis}

The bond operators ${\hat A}_{ij}$ and ${\hat B}_{ij}$ are both invariant under global SU(2) spin rotations
${\cal U}_{\bm n}(\varphi)$ in the original reference frame. In the twisted reference frame, this SU(2) symmetry group 
is generated by \textit{staggered} U(1) spin rotations, ${\cal U}^{\prime}_{\bm x}$ and  ${\cal U}^{\prime}_{\bm z}$, about the $x$ and $z$ axes, and by
a \textit{uniform} U(1) spin rotation ${\cal U}^{\prime}_{\bm y}$ about the $y$ axis. The transformation rules of the  Schwinger bosons under these symmetry operations read
\begin{align}
{\cal U}^{\prime}_{\bm x} (\varphi) {\hat {\bm b}}_{j} {\cal U}^{\prime \dagger}_{\bm x} (\varphi) &= e^{-i\eta_j {\varphi \over 2} \sigma_x} {\hat {\bm b}}_j, \label{eq:sx} \\
{\cal U}^{\prime}_{\bm y} (\varphi) {\hat {\bm b}}_{j} {\cal U}^{\prime \dagger}_{\bm y} (\varphi) &= e^{-i{\varphi \over 2} \sigma_y} {\hat{\bm b}}_j, \label{eq:sy} \\
{\cal U}^{\prime}_{\bm z} (\varphi) {\hat {\bm b}}_{j} {\cal U}^{\prime \dagger}_{\bm z} (\varphi)&= e^{-i\eta_j {\varphi \over 2} \sigma_z} {\hat {\bm b}}_j,  \label{eq:sz}
\end{align}
where $\eta_j=+1$ ($-1$) on the $A$ ($B$) sublattice.
In momentum space, these transformations become
\begin{align}
{\cal U}^{\prime}_{\bm x} (\varphi) {\hat {\bm b}}_{\bm q} {\cal U}^{\prime \dagger}_{\bm x} (\varphi)  &= \cos {\varphi \over 2} {\hat {\bm b}}_{\bm q} - i \sin {\varphi \over 2} \sigma_x {\hat {\bm b}}_{{\bm q} + {\bm \pi}}, \label{eq:sxq} \\
{\cal U}^{\prime}_{\bm y} (\varphi) {\hat {\bm b}}_{\bm q} {\cal U}^{\prime \dagger}_{\bm y} (\varphi) &= e^{-i{\varphi \over 2} \sigma_y} {\hat {\bm b}}_{\bm q}, \label{eq:syq} \\
{\cal U}^{\prime}_{\bm z} (\varphi) {\hat {\bm b}}_{\bm q} {\cal U}^{\prime \dagger}_{\bm z} (\varphi)  &= \cos {\varphi \over 2} {\hat {\bm b}}_{\bm q} - i \sin {\varphi \over 2} \sigma_z {\hat {\bm b}}_{{\bm q} + {\bm \pi}}. 
\label{eq:szq}
\end{align}
The SP Hamiltonian remains invariant under these  transformations because the bond operators ${\hat A}_{ij}$ and ${\hat  B}_{ij}$ are SU(2) invariants. 
By taking $\varphi=\pi$, we obtain
\begin{align}
w^{\dagger}_{\mu} {H}_{\rm sp}(\bm q)  w_{\mu} &= H_{\rm sp}({\bm q} + {\bm \pi}), \;\; {\rm for} \;\; \mu=x,z, \label{wxz} \\
w^{\dagger}_{y} H_{\rm sp}(\bm q) w_{y} &= H_{\rm sp}({\bm q}),\label{wy}
\end{align}
where $w_{\mu} = -i \sigma_z \otimes \sigma_{\mu}$ and $w_{y} = -i \sigma_0 \otimes \sigma_{y}$ are the matrices associated with the $\pi$-rotations of the Nambu spinor Eq.~\eqref{fns} along the $\mu$ axis.
After the para-unitary diagonalization \cite{Colpa1978}, the spectrum of the SP Hamiltonian is determined by
the eigenvalue equations:
\begin{eqnarray}
\label{SP_solution}
\gamma^0 H_{\rm sp}(\bm{q})X_{\bm{q} \sigma} &=& \varepsilon_{\bm{q} \sigma} X_{\bm{q} \sigma}, 
\nonumber \\
\gamma^0 H_{\rm sp}(\bm{q})\bar{X}_{\bm{q} \sigma} &=&-\varepsilon_{-\bm{q} \sigma}\bar{X}_{\bm{q} \sigma}.
\end{eqnarray}
Eq.~(\ref{wxz}) implies  that $\varepsilon_{{\bm q} \sigma}=\varepsilon_{{\bm q} + {\bm \pi} \sigma}$ and $X_{\bm{q} \sigma} = w_{x/z} X_{\bm{q}+{\bm \pi} \sigma}$.
In addition, Eq.~(\ref{wy}) implies that the eigenstates of $H_{sp}({\bm q})$ are also eigenstates of $w_y$: $w_y X_{\bm{q} \sigma} = \pm  X_{\bm{q} \sigma}$ because $w_y^2 = -I$. 
The same symmetry argument holds for the eigenstates with negative eigenvalues.

While a non-zero expectation value of the bond operator ${\hat B}_{\bm \delta}$ is allowed by the symmetries of the SP Hamiltonian,
this expectation value turns out to be zero because of a cancellation between matrix elements. For instance, in the SP approximation, 
\begin{eqnarray}\label{eq:Bx_expectation}
B_x &=& (1/2) \sum_{\sigma {\bm q}} [ \bar{X}^{\dagger}_{{\bm q} \sigma} iw_y \bar{X}_{ {\bm q}\sigma} (1+n_B(\varepsilon_{{\bm q} \sigma})) 
\nonumber \\
&+& {X}^{\dagger}_{{\bm q}\sigma} iw_y {X}_{{\bm q}\sigma} n_B(\varepsilon_{{\bm q} \sigma})], 
\end{eqnarray}
where $n_B(\varepsilon)$ is the Bose distribution function. Since $\{w_y,w_z\}=0$,
$\bar{X}_{ {\bm q} \sigma}$ and $\bar{X}_{{\bm q}+{\bm \pi} \sigma}$, or ${X}_{ {\bm k} \sigma}$ and ${X}_{ {\bm q}+{\bm \pi} \sigma}$, must be the eigenvectors of $w_y$ with opposite eigenvalues. The relation $X_{\bm{q} \sigma} = w_{x/z} X_{\bm{q}+{\bm \pi} \sigma}$ leads to $\bar{X}^{\dagger}_{{\bm q} \sigma} iw_y \bar{X}_{ {\bm q} \sigma} = - \bar{X}^{\dagger}_{{\bm q}+{\bm \pi} \sigma} iw_y \bar{X}_{{\bm q}+{\bm \pi} \sigma}$ and ${X}^{\dagger}_{{\bm q} \sigma} iw_y {X}_{ {\bm q} \sigma}=-{X}^{\dagger}_{{\bm q}+{\bm \pi} \sigma} iw_y {X}_{ {\bm q}+{\bm \pi} \sigma}$. Since the occupation numbers $n_B(\varepsilon_{{\bm q} \sigma})$ at ${\bm q}$ and ${\bm q+\pi}$ are the same because $\varepsilon_{{\bm q}\sigma}=\varepsilon_{{\bm q} + {\bm \pi} \sigma}$, we obtain $B_x =0 $ at any finite temperature $T$.
As we will show later, at $T=0$ spinons can condense in multiple ways  with different macroscopic occupation numbers of the single-spinon ${\bm q} = {\bm 0}\equiv(0,0)$ and ${\bm \pi}\equiv(\pi,\pi)$ states. 
Nevertheless, we have verified that  $B_x$ remains equal to zero for any of these alternative SP solutions. 
Consequently, the general SP solutions that we will consider here are described by the bond field ${\hat A}_{\bm \delta}$, which  is invariant  under a staggered gauge transformation 
\begin{align}
C(\varphi) {\hat {\bm b}}_{j} C^{\dagger}(\varphi) &= e^{-i \eta_j {\varphi \over 2}} {\hat {\bm b}}_{j}, \label{eq:u1c}
\end{align}
or, in momentum space,
 \begin{align}
C(\varphi) {\hat {\bm b}}_{\bm q} C^{\dagger}(\varphi) &= \cos {\varphi \over 2} {\hat {\bm b}}_{\bm q} - i \sin {\varphi \over 2} {\hat {\bm b}}_{{\bm q} + {\bm \pi}}. \label{eq:u1c_q}
\end{align}

The Hamiltonian invariance under gauge transformations restricts the form of the momentum space version of the SP Hamiltonian (see Eq.~\eqref{Hspq}). Under the gauge transformation $C(\varphi=\pi)$, the four dimensional spinor $\psi({{\bm q}})$ is transformed into 
$w_0 \psi({{\bm q} + \bm \pi})$ with $w_0 = -i \sigma_z \otimes \sigma_0$. 
Since $w_0^{\dagger}H_{\rm sp}({\bm q})w_0 = H_{\rm sp}({\bm q}+{\bm \pi})$, we conclude that 
$\varepsilon_{{\bm q} \sigma}=\varepsilon_{{\bm q} + {\bm \pi} \sigma}$ and $X_{\bm{q} \sigma} = w_{0} X_{\bm{q}+{\bm \pi} \sigma}$.

There is also a ``particle-hole'' symmetry $PH_{\rm sp}^*({\bm q})P = H_{\rm sp}(-{\bm q})$ inherited from the Nambu representation. 
Given that $\{P,\gamma^0\}=0$, Eq.~\eqref{SP_solution} implies that $P {\bar X}_{ {\bm q} \sigma}^*$ and $P X_{{\bm q} \sigma}^*$ must be the eigenvectors of $H_{\rm sp}(-{\bm q})$ with eigenenergies $\varepsilon_{-{\bm q} \sigma}$ and $-\varepsilon_{{\bm q} \sigma}$, respectively. 
In the presence of the inversion symmetry, we also have $\varepsilon_{-{\bm q} \sigma}=\varepsilon_{{\bm q} \sigma}$.

\subsection{Single-spinon spectrum}

Since  ${\cal B}_{\bm \delta}=0$ at the SP level, the SP SU(2) Hamiltonian has two degenerate single-spinon   energy branches 
\begin{align}
\varepsilon_{\bm{q} \sigma} \equiv \varepsilon_{{\bm q}} = \sqrt{ \lambda^{2}-\Delta_{\bm{q}}^{2}}. 
\label{e_spinon}
\end{align}
$\sigma= \pm1$ represents the spin quantum number, $\uparrow,\downarrow$, of these spinon modes,
whose eigenvectors  are
\begin{equation} \label{eigenstate1}
{X_{\bm{q},+1}}=\left(\begin{array}{c}
u_{\bm{q}}\\
0\\
v_{ \bm{q}}\\
0
\end{array}\right), \;\;\; 
{X_{\bm{q},-1}}=\left(\begin{array}{c}
0\\
u_{\bm{q}}\\
0\\
v_{\bm{q}}
\end{array}\right),
\end{equation}
with
\begin{align}\label{coherent_factor}
u_{\bm{q}} & =\sqrt{\frac{1}{2}\left(\frac{\lambda  }{\varepsilon_{\bm{q}}}+1\right)},\;
v_{\bm{q}}=-\frac{\Delta_{\bm{q}}}{\rvert\Delta_{\bm{q}}\rvert}\sqrt{\frac{1}{2}\left(\frac{\lambda }{\varepsilon_{\bm{q}}}-1\right)}.
\end{align}
The corresponding eigenvectors  for negative energy eigenvalues, $-\varepsilon_{{\bm q}}$, are
\begin{equation}\label{eigenstate2}
{\bar{X}_{\bm{q},+1}}=\left(\begin{array}{c}
v_{\bm{q}}\\
0\\
u_{\bm{q}}\\
0
\end{array}\right),\; \; \;
{\bar{X}_{\bm{q},-1}}=\left(\begin{array}{c}
0\\
v_{\bm{q}}\\
0\\
u_{\bm{q}}
\end{array}\right).
\end{equation}

In accordance with Goldstone's theorem, the spinon condensation  leads to a gapless spinon dispersion $\varepsilon_{\bm{q}\sigma} \equiv \varepsilon_{{\bm q}} $ in the thermodynamic limit.
The gapless modes have momenta
${\bm q}={\bm 0}$ and ${\bm \pi}$, implying that $\lambda = 2(1-\alpha)J A_x $.
The gauge freedom of the theory, $b_{i\sigma}(\tau) \rightarrow b_{i\sigma}(\tau) e^{i \theta_{i}(\tau)}$, allows us  to assume that $A_x >0$. An explicit solution of the SP equations [Eqs.~\eqref{self-sp-eq1} to \eqref{self-sp-eq3}] gives at $T=0$
\begin{align}
A_x  = 2S+c_1,   \;\;\; B_x=0,   \;\;\;
n_c= 2S - c_2,
\end{align}
where
\begin{align}
c_1 &= \int {d^2 {\bm k} \over (2\pi)^2} \left( 1-\sqrt{1-(\gamma^+_{\bm k})^2} \right) \approx 0.158, \\
c_2 &= \int {d^2 {\bm k} \over (2\pi)^2} \left( {1\over \sqrt{1-(\gamma^+_{\bm k})^2}} - 1 \right) \approx 0.393,
\end{align}
and $n_c$ is the spinon  condensate fraction.
Since $n_c$ must be positive, the above solution is valid only for $2S>0.393$, which includes all the physical values of $S$.

\subsection{Spinon condensation}

\begin{figure}[!t]
\centering
\includegraphics[scale=0.5]{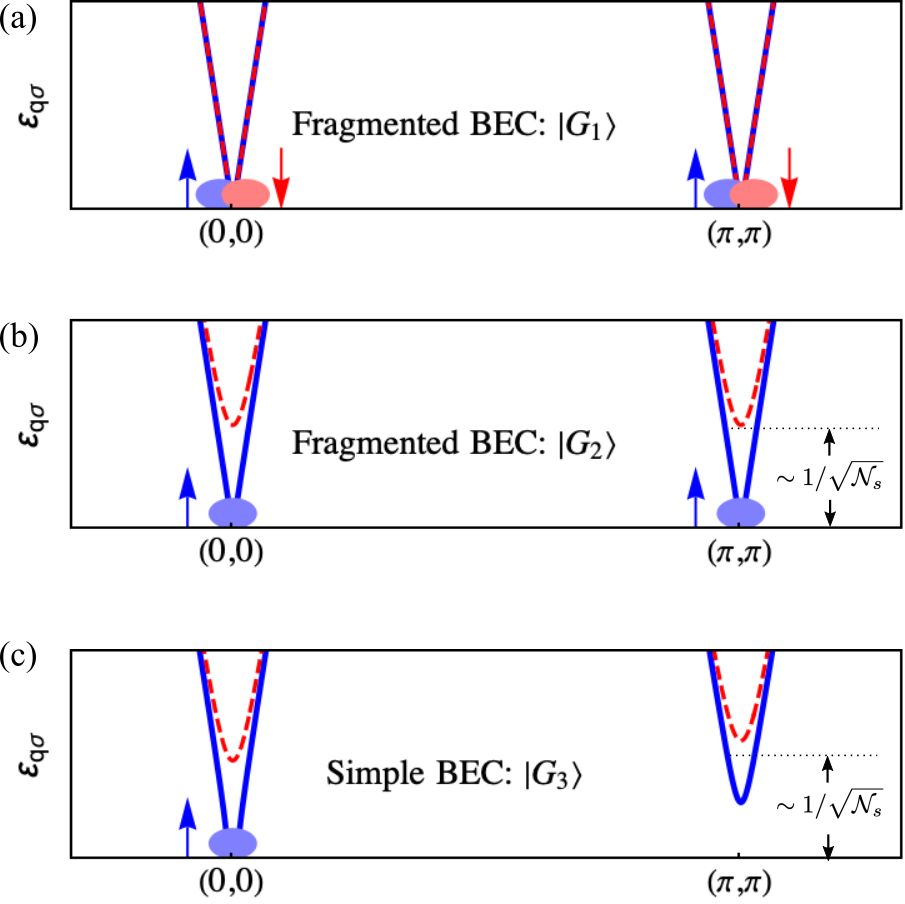}
\caption{Single-spinon spectrum of square lattice antiferromagnet  in absence of a symmetry-breaking field (a), in the presence of a symmetry-breaking field $h$ linearly coupled with the staggered magnetization (b) and 
with an additional symmetry-breaking field $t$ that couples to a single-spinon nearest-neighbor hopping (c), where $h,t \sim 1/{\cal N}_s$.}
\label{fig:condensate}
\end{figure}

\subsubsection{Fragmented versus simple BEC}

The above SP Hamiltonian has four zero-energy single-spinon states with ${\bm q}={\bm 0} $ or  ${\bm \pi}$ and $\sigma=\pm1$ in the thermodynamic limit, implying that  the spinons can condense in multiple ways at zero temperature.
We note, however, that this degeneracy is not present on finite size lattices because the single-spinon spectrum has a finite size gap that forces the SP solution to be unique and SU(2) invariant. Consequently, in the absence of any external symmetry breaking field, the ground state of the SP Hamiltonian remains SU(2) invariant upon taking the thermodynamic limit, implying that
spinons condense  on the four zero-energy modes with equal condensate fractions:
\begin{eqnarray}
\rvert G_1\rangle &=& 
\frac{1}{u_{\bm{0}  \uparrow}^{4}}
e^{-\frac{v_{\bm{0} \uparrow}^{*}}{2 u_{\bm{0} \uparrow}^{*}} \sum_{\sigma} (\hat{b}_{\bm{0}\sigma}^{\dagger} \hat{b}_{\bm{0}\sigma}^{\dagger}- \hat{b}_{\bm{\pi} \sigma}^{\dagger} \hat{b}_{\bm{\pi} \sigma}^{\dagger})}
\nonumber \\
&\times & \prod_{{\bm k}\neq {\bm 0},{\bm \pi},\sigma} \frac{1}{u_{\bm{k} \sigma}}e^{-\frac{v_{\bm{k} \sigma}^{*}}{2 u_{\bm{k \sigma}}^{*}} \hat{b}_{\bm{k}\sigma}^{\dagger}\hat{b}_{-\bm{k}\sigma}^{\dagger}}\rvert\emptyset\rangle,
\label{su2bec}
\end{eqnarray}
 where $\rvert \emptyset \rangle$ is the vacuum of the Schwinger bosons $b_{i\sigma}$
(see Fig.~\ref{fig:condensate} (a)).~\footnote{We note that we are choosing condensate states with well-defined parity of the number of SB's. Since the ground states in the even and odd parity sectors become degenerate in the thermodynamic limit, it is also possible to choose condensates without a well-defined parity that lead to $ \langle b_{j \sigma} \rangle \neq 0$. Given that this choice does not modify the expectation values of physical observables, here we work with condensates with well-defined parity for simplicity.}
The average occupation number $n_{\bm k}$ of Schwinger bosons with momentum ${\bm k}$ is determined by the ratio
$\rvert v_{{\bm k} \sigma}/u_{\bm k \sigma} \rvert = \sqrt{ (\lambda- \varepsilon_{{\bm k} \sigma} )/(\lambda  +\varepsilon_{{\bm k} \sigma})} $. 
This ratio satisfies that $\lim_{{\cal N}_s \to \infty} \rvert v_{{\bm k} \sigma}/u_{\bm k \sigma} \rvert \to 1$ for ${\bm k}={\bm 0}, {\bm \pi}$, i.e., at the wave vectors where the spinon spectrum is gapless.
The finite size scaling satisfies  $1 - \rvert v_{\bm{0} \uparrow}/u_{\bm{0} \uparrow}\rvert  \propto 1/{\cal N}_s$ because  the finite size gap scales as $1/{\cal N}_s$, $\varepsilon_{{\bm 0} \uparrow} \propto 1/{\cal N}_s$, implying a macroscopic occupation, $\propto {\cal N}_s$, of the four zero-energy single-spinon states.
As we will discuss below, this SU(2) invariant spinon condensate corresponds to a ``fragmented'' BEC.~\cite{penrose1956bose,leggett2006quantum}

As usual, to account for the  N{\' e}el antiferromagnetic order, we must introduce an external symmetry breaking field $h$ linearly coupled to the  staggered magnetization.  In the twisted reference frame, this coupling takes the form  $- h \sum_{i, \sigma} \sigma \hat{b}^{\dagger}_{i\sigma}\hat{b}_{i\sigma}^{} $ and the field $h$ is sent to zero after taking the thermodynamic limit. 
Since the symmetry breaking field $h$ gaps out the single-spinon branch with spin index $\downarrow$ (the  field $h$ is applied along the positive $z$-direction in the twisted reference frame),
the spinons must condense in the remaining two zero-energy states with spin index $\uparrow$, implying that the ground state degeneracy is not completely removed [see Fig.~\ref{fig:condensate} (b)]. 
By analogy with the SU(2) invariant solution \eqref{su2bec}, the bosons can still condense in a fragmented BEC 
\begin{eqnarray}
\rvert G_2\rangle &=& 
\frac{1}{u_{\bm{0} \uparrow}^{2}}e^{-\frac{v_{\bm{0} \uparrow}^{*}}{2 u_{ \bm{0} \uparrow}^{*}} (\hat{b}_{\bm{0\uparrow}}^{\dagger} \hat{b}_{\bm{0}\uparrow}^{\dagger} 
- \hat{b}_{\bm{\pi} \uparrow}^{\dagger}  \hat{b}_{\bm{\pi} \uparrow }^{\dagger})}
\nonumber \\ 
&\times &  \!\!\!\!\!\!\!\!\!\!\!\! \!\!\!\! 
\prod_{({\bm k},\sigma)\neq ({\bm 0},\uparrow),({\bm \pi},\uparrow)} 
\frac{1}{u_{\bm{k} \sigma }}e^{-\frac{v_{\bm{k} \sigma}^{*}}{2 u_{\bm{k} \sigma}^{*}} \hat{b}_{\bm{k}\sigma}^{\dagger}\hat{b}_{-\bm{k}\sigma}^{\dagger} } 
\rvert\emptyset\rangle,
\label{G2}
\end{eqnarray}
where $1-\rvert v_{\bm{0} \uparrow}/u_{\bm{0} \uparrow} \rvert \propto  1/{\cal N}_s$ and $1-\rvert v_{\bm{0} \downarrow}/u_{\bm{0} \downarrow} \rvert \propto  1/{\cal N}_s^{\nu}$, where $\nu <1$ is controlled by the finite size scaling of the symmetry breaking field. For instance,  the finite size gaps for spin up and spin down spinons are  $\varepsilon_{ {\bm 0} \uparrow} = \varepsilon_{ {\bm \pi} \uparrow} \propto 1/{\cal N}_s$ and $\varepsilon_{{\bm 0}\downarrow} = \varepsilon_{{\bm \pi}\downarrow} \propto 1/{{\cal N}_s^{1/2}}$ for a symmetry-breaking field $h \propto 1/{\cal N}_s$. 
In  other words, the  BEC state Eq.~\eqref{G2} describes a  macroscopic occupation of both zero-energy states with spin index up.

We note, however, that previous SB approaches to the square lattice AFM Heisenberg have always adopted the SP solution $\rvert G_2\rangle$ without an explicit justification~\cite{book2012interacting}. In fact, at the SP level, the magnetically ordered state can be equally described by an infinite number of \emph{fragmented} spin condensates with different condensate fractions $(n_{\bm 0},n_{\bm \pi})=(n_1,n_c-n_1)$ at the two orthogonal zero energy modes
with $0 < n_1 < n_c$. These condensates 
are called the ``fragmented''  because the single-particle density matrix has two macroscopic eigenvalues
${\cal N}_s n_1$ and ${\cal N}_s (n_c-n_1)$. 
We note that the choice of the  two zero energy modes with different condensate fractions is made by fixing  the gauge transformation $C(\varphi)$ introduced in Eq.~\eqref{eq:u1c_q}. In particular, the ground state $\rvert G_2\rangle$ corresponds to the gauge choice $\varphi=0$ and $n_{ {\bm 0}}=n_{ {\bm \pi}}=n_c/2$. Because of this gauge freedom, 
the condensate fractions $(n_1,n_2)$ and $(n_2,n_1)$ describe the same physical state because the two zero modes can be exchanged by applying the gauge transformation $C(\pi)$ [see Eq.~~\eqref{eq:u1c_q}]

The ground state of the mean field Hamiltonian with $n_{ 1} = n_c$, or the gauge equivalent one with $n_{2}=n_c$, is called ``simple'' BEC because the single-particle density matrix has only one macroscopic eigenvalue~\cite{leggett2006quantum}. We note that there is only one \emph{physical} simple BEC ground state. 
In other words, all bosons are condensed in the single mode $b_1^\dag = \cos{\frac{\varphi}{2}} \ b_{\bm 0, \uparrow}^\dag + i \sin{\frac{\varphi}{2}} \ b_{\bm \pi, \uparrow}^\dag $, where the continuous variable $\varphi$ parametrizes the gauge orbit of physically equivalent simple BEC solutions. 
This simple BEC solution, for a particular gauge choice, can be selected by adding the infinitesimal ``symmetry breaking  field'' term,
\begin{equation*}
    t\sum_{\langle ij \rangle, \sigma} (C_{ij}^{\dag}(\varphi) + C_{ij}^{}(\varphi))
\end{equation*}
to the mean field Hamiltonian  with
\begin{equation}
    C_{ij}^\dag(\varphi) = \frac{1}{2} \bm b_i^\dag ( \bm \sigma \cdot \bm n(\varphi) ) \bm b_j^{}  \nonumber,
\end{equation}
in the global reference frame and  $\bm n(\varphi) = (\cos \varphi, \sin \varphi, 0)$. We note that this operator transforms like a triplet under SU(2) rotations and breaks the U(1) symmetry of global rotations about the $z$-axis.
In the twisted reference frame, the triplet bond operator $C_{ij}^\dag(\varphi)$ becomes
\begin{equation}
    C_{ij}^\dag(\varphi) = -\frac{1}{2} \sum_{\sigma} e^{- i\sigma \eta_i \varphi} \sigma b_{i\sigma}^\dag b_{j \sigma}^{}.   \nonumber
\end{equation}

For the gauge choice $\varphi=0$, the simple BEC state $\rvert G_3\rangle$ can be expressed as
\begin{eqnarray}
\rvert G_3\rangle =
\frac{1}{u_{\bm{0} \uparrow}}e^{-\frac{v_{\bm{0} \uparrow}^{*}}{2 u_{\bm{0} \uparrow}^{*}} \hat{b}_{\bm{0}\uparrow}^{\dagger}\hat{b}_{\bm{0}\uparrow}^{\dagger}} \!\!\!\!\!\!\!\!\!
\prod_{({\bm k},\sigma)\neq ({\bm 0},\uparrow)} \frac{1}{u_{ \bm{k} \sigma}}e^{-\frac{v_{ \bm{k} \sigma}^{*}}{2 u_{\bm{k} \sigma }^{*}}\hat{b}_{\bm{k}\sigma}^{\dagger}\hat{b}_{-\bm{k}\sigma}^{\dagger} } 
\rvert\emptyset\rangle,
\end{eqnarray}
where, assuming $h,t\sim 1/{\cal N}_s$, the ratio $\rvert v_{\bm{Q} \sigma}/u_{\bm{Q} \sigma} \rvert$ satisfies $1-\rvert v_{\bm{Q} \sigma}/u_{\bm{Q} \sigma} \rvert \propto  1/{\cal N}_s$ for ${\bm Q}={\bm 0}$ and $\sigma=\uparrow$ and  $ 1- \rvert v_{\bm{Q} \sigma}/u_{\bm{Q} \sigma} \rvert \propto  1/{\cal N}_s^{1/2}$  for the other three zero-energy modes.  In other words, $\rvert G_3\rangle$ describes a BEC with \emph{macroscopic occupation of the single mode} ${\bm k}={\bm 0}$ and $\sigma=\uparrow$  [see Fig.~\ref{fig:condensate} (c)].

As we will see in the next subsection, while the ground states $\rvert G_2\rangle$ and $\rvert G_3\rangle$ correspond to   ferromagnetic SP solutions in the twisted reference frame, they do not represent the same physical state because 
$\langle  G_3 \rvert  {\hat O} \rvert G_3\rangle \neq \langle  G_2 \rvert  {\hat O} \rvert G_2\rangle$ for some physical observables. Although these different condensates are degenerate at the SP level,
 fragmented BECs are known to be ``fragile'' against the inclusion of boson-boson interactions, which typically favor the simple BEC state~\cite{penrose1956bose,leggett2006quantum}. This simple observation suggests that fluctuations of the HS fields should select the SP solution $\rvert G_3\rangle$ corresponding to the simple BEC. Indeed, as we will see in Sec.~\ref{MP}, the expansion around the SP solution represented by the ground state $\rvert G_2\rangle$ 
 leads to a dynamical spin structure factor that is qualitatively incorrect for $\alpha \neq 0$ 
 (the linear spin wave result is not recovered in the large-$S$ limit). In contrast, the large-$S$ limit of the dynamical spin structure factor that is obtained by including  fluctuations around the simple BEC SP solution represented by the state $\rvert G_3\rangle$  coincides with the linear spin wave theory result.

\subsubsection{Physical character of candidate ground states}

The invariance of  $\rvert G_1\rangle$ under global SU(2) spin rotations in the original reference frame implies that the local magnetization is zero for this SP solution. In contrast, the  ground state $\rvert G_{2}\rangle$ is U(1) invariant under global (staggered) spin rotations about the $z$ axis in the original (twisted) reference frame, which  is compatible with a finite local magnetization ${\bm m}= m {\hat {\bm z}}$ with  $m=n_c/2$. The ground state $\rvert G_{3}\rangle$ is invariant under a composition of the same U(1) spin rotation by an angle $\varphi$
about the $z$ axis and the staggered gauge transformation $C(-\varphi)$ introduced in Eq.~\eqref{eq:u1c_q}. In other words, ${\tilde {\cal U}}_{\bm z}(\varphi) \rvert G_{3}\rangle$ 
[$ {\tilde {\cal U}}_{\bm z}(\varphi)\equiv {\cal U}_{\bm z}(\varphi) C(-\varphi) $]
generates  an orbit of gauge equivalent states. Although the local magnetization is also  ${\bm m}= m {\hat {\bm z}}$ with  $m=n_c/2$, the states  
$\rvert G_{2}\rangle$ and $\rvert G_{3}\rangle$ are not connected by a gauge transformation. For instance, the two states have  different expectation values of on-site spin correlation functions:
\begin{eqnarray}
\langle   G_{2} \rvert \hat{\vec{S}}_{i}^{\bot} \cdot \hat{\vec{S}}_{i}^{\bot} \rvert G_{2}\rangle &=&  S(S+1)-m^2,
\\ \nonumber
\langle   G_{3} \rvert \hat{\vec{S}}_{i}^{\bot} \cdot \hat{\vec{S}}_{i}^{\bot} \rvert G_{3}\rangle &=& S(S+1)-m^2,
\\ \nonumber 
\langle   G_{2} \rvert \hat{S}_{i}^{z}\hat{S}_{i}^{z}  \rvert G_{2}\rangle &=& \frac{1}{2}S(S+1)+\frac{1}{2} m^{2},
\\ \nonumber 
\langle   G_{3} \rvert  \hat{S}_{i}^{z}\hat{S}_{i}^{z} \rvert G_{3}\rangle &=& \frac{1}{2}S(S+1)+\frac{3}{2} m^{2}, 
\end{eqnarray}
where $\hat{\vec{S}}_{i}^{\bot}=(\hat{S}_i^{x}, \hat{S}_i^y)$ are the transverse components of the spin operator relative to the magnetization axis. We note that these expectation values are independent of $\alpha$. 
The resulting expectation value of $\hat{{\bm S}}_i \cdot \hat{{\bm S}}_i$ that determines the sum rule of the dynamical spin structure factor \emph{at the SP level} is:
\begin{eqnarray}
\langle   G_{2} \rvert  \hat{{\bm S}}_i \cdot \hat{{\bm S}}_i \rvert G_{2}\rangle &=& 3 S(S+1)/2 - m^2/2
\nonumber \\
\langle   G_{3} \rvert \hat{\bm S}_i \cdot \hat{\bm S}_i  \rvert G_{3}\rangle &=& 3 S(S+1)/2 + m^2/2
\label{srsp}
\end{eqnarray}
Both results are different from the one obtained for the SU(2) invariant condensate: 
$\langle   G_{1} \rvert  \hat{\bm S}_i \cdot \hat{\bm S}_i \rvert G_{1}\rangle = 3 S(S+1)/2$.   Clearly, the three condensates violate the  identity $\hat{\bm S}_i \cdot \hat{\bm S}_i= S(S+1)$ [Casimir operator of SU(2)] because the local constraint Eq.~\eqref{constraint} is only imposed at the mean field level. 
This shortcoming of the SP approximation leads to a well-known violation of the sum rule of the dynamical spin structure factor because its integral over momentum and frequency is equal to $\langle  \hat{\bm S}_i \cdot \hat{\bm S}_i \rangle$~\cite{book2012interacting}. In other words, according to Eqs.~\eqref{srsp}, the frequency and momentum integral of the spectral weight is overestimated by approximately 50\% [exactly $50\%$ for the SU(2) invariant condensate]. As we will see in Sec.~\ref{MP},  the sum rule is recovered to a much better approximation upon including fluctuations around the SP solution $\rvert G_{3}\rangle$ and this result remains basically independent of $\alpha$.

Hereafter, we will consider simple BEC SP solution $\rvert G_3 \rangle $ for reasons that will become clear upon including 
$1/N$ corrections. 
For concreteness, we will adopt the gauge in which  the bosons condense at the single-spinon ${\bm Q}={\bm 0}$ state, which is invariant under spatial inversion.

For later use, we introduce the  ``bare'' or SP  single-spinon Green's function that is obtained with the simple condensate SP solution $\rvert G_3\rangle$:
\begin{eqnarray}\label{green}
{\cal G}_{\rm SP}(\bm{q},i\omega_n) = {\cal G}_{n}(\bm{q},i\omega_n) + {\cal G}_{c}(\bm{q},i\omega_n).
\end{eqnarray}
The first term describes the non-condensed spinon at $(\bm{q},\sigma) \neq (\bm{0},1)$, which is equal to the
inverse of the dynamical matrix
\begin{eqnarray}\label{green_n}
 {\cal G}_{n}(q)  =  [2{\cal M}(q)]^{-1}  =  \sum_{\sigma=\pm 1} 
\frac{g_{\bm{q} \sigma }^{-}}{\varepsilon_{\bm{q} \sigma }-i\omega_{n}}+\frac{g_{\bm{q} \sigma }^{+}}{\varepsilon_{-\bm{q}\sigma}+i\omega_{n}},\label{eq:GF}
\end{eqnarray}
where $g_{\bm{q}\sigma}^{-}=X_{\bm{q}\sigma}X_{\bm{q}\sigma}^{\dagger}$ and $g_{\bm{q}\sigma}^{+}=\bar{X}_{\bm{q}\sigma}\bar{X}_{\bm{q}\sigma}^{\dagger}$.
The second term arises from the condensed boson
\begin{align} 
{\cal G}_{c}(\bm{q},i\omega_{n}) &= {\cal N}_{s} g_{c}^{{\bm Q}}  \delta_{{\bm q},{\bm Q}} \left( {1\over \epsilon_c - i\omega_n} + {1\over \epsilon_c + i\omega_n} \right) \nonumber \\
& \stackrel{{\cal N}_{s} \to \infty}{=} (2\pi)^3 g_{c}^{{\bm Q}} \delta({\bm q}-{\bm Q}) \delta(\omega_n),
\label{green_c}
\end{align}
where ${\bm Q}={\bm 0}$, $g_{c}^{{\bm Q}} =  \rvert c_{{\bm Q}} \rangle \langle  c_{{\bm Q}} \rvert$ with $\rvert c_{\bm{Q}} \rangle \equiv \lim_{{\cal N}_s \rightarrow \infty} X_{{\bm Q},+1}/\sqrt{{\cal N}_s}  = \sqrt{n_c}(1,0,-1,0)^T$.
The second line of Eq.~\eqref{green_c} corresponds to  the thermodynamic limit in which the finite size gap of the 
single-spinon eigenstate with ${\bm Q}={\bm 0}$ and $\sigma  =+1$  goes to zero: $\epsilon_c \rightarrow 0^+$.

\subsection{Dynamical spin structure factor}
\subsubsection{General formulation}
Our next goal is to compute the DSSF in the SP approximation. 
We will present the result for arbitrary values of $N$ because in a later section we will consider higher order corrections in powers of $1/N$. The first step is to introduce a Zeeman coupling to an external time and space dependent magnetic field, i.e., to add source terms to the action 
\begin{eqnarray}
N S_{s} = - {1\over 2} \int _0^{\beta} d\tau \sum_{j \mu} h_j^{\mu}(\tau) {\psi}^{\dagger}_j (\tau) u^{\mu} \psi_j(\tau),
\end{eqnarray}
where ${1\over 2}  {\psi}^{\dagger}_j u^{\mu} \psi_j \equiv S_j^{\mu}$ represents the $\mu$ component of the spin $j$ (note that we are using a single index $\mu$ to denote the generators of $Sp(N/2)$). 
This source term modifies the dynamical matrix to ${\cal M}_h(k; p) = {\cal M}(k; p) -  1/(\sqrt{{\cal N}_s \beta}) h_{k-p}^{\mu} u^{\mu}$, whose first derivative
with respect to the field $h_{k-p}^{\mu}$ gives rise to the external vertices
\begin{eqnarray}
u^{\mu} { (k, p) } =  - \left(\frac{1}{ \sqrt{{\cal N}_{s}\beta}}\right)^{-1} \frac{\delta {\cal M}_h(k; p)}{\delta h_{k-p}^{\mu} }.\label{Sext}
\end{eqnarray}

The dynamical magnetic susceptibility is then given by
\begin{eqnarray}
\chi^{\mu\nu}(q) & = & {\delta^2 {\ln{{\cal Z}}} \over \delta h^{\mu}_{-q} \delta h^{\nu}_{q}},
\end{eqnarray}
where $h^{\mu}_q$ is the Fourier component of the external fields $h^{\mu}_i(\tau)$. 
According to the fluctuation-dissipation theorem~\cite{coleman2015introduction}, the DSSF is related to the imaginary part 
of the magnetic susceptibility. At zero temperature we have:
\begin{equation}
S^{\mu\nu}({\bm q},\omega) = \frac{1}{\pi} \Theta(\omega) \text{Im} [ \chi^{\mu\nu}({\bm q},\omega) ],
\end{equation}
with $\Theta(\omega)$ being the Heaviside step function.

\begin{figure}[!t]
\centering
\includegraphics[scale=0.6]{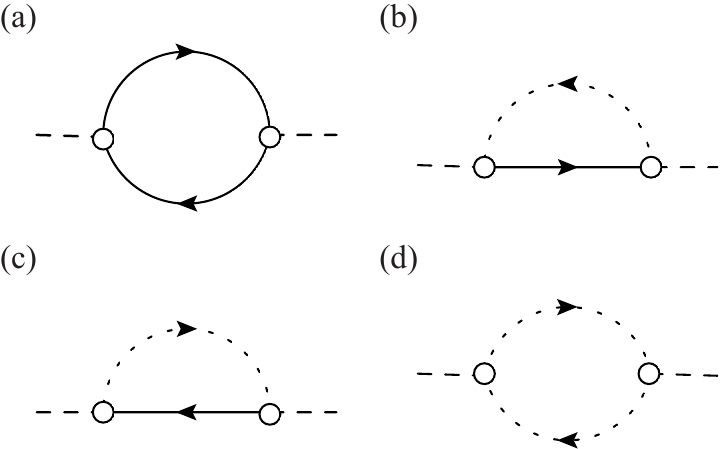}
\caption{Diagramatic representation of the four contributions to the dynamical spin susceptibility $\chi^{\mu\nu}_{\rm SP}(q)$ at the SP level corresponding to the four terms of   Eq.~\eqref{eq:chisp}). The full (dashed) line represents the propagator of the non-condensed (condensed)  spinons.  The open circles represent the external vertices of the theory introduced in Eq.~(\ref{Sext}).  }
\label{fig:chisp}
\end{figure}

At the SP level, the dynamical magnetic susceptibility is obtained by taking the second functional derivative 
of   ${\ln ({\cal Z}_{\rm SP})}$:
\begin{align}\label{eq:chisp}
\chi^{\mu\nu}_{\rm SP}(q) & =  {1\over {\cal N}_s \beta} \sum_{k} \frac{1}{2} \text{Tr}[{\cal G}_n(k) u^{\mu}{\cal G}_n(k+q)u^{\nu}]  \nonumber \\
&+{1\over {\cal N}_s \beta} \sum_{k} \frac{1}{2} \text{Tr}[{\cal G}_c(k) u^{\mu}{\cal G}_n(k+q)u^{\nu}]  \nonumber \\
&+{1\over {\cal N}_s \beta} \sum_{k} \frac{1}{2} \text{Tr}[{\cal G}_n(k) u^{\mu}{\cal G}_c(k+q)u^{\nu}]  \nonumber \\
&+{1\over {\cal N}_s \beta} \sum_{k} \frac{1}{2} \text{Tr}[{\cal G}_c(k) u^{\mu}{\cal G}_c(k+q)u^{\nu}],
\end{align}
which can be decomposed into four contributions represented by   the four diagrams (a)-(d) in Fig.~\ref{fig:chisp}.
The full lines represent the normal contribution to the Green's functions, while the dashed lines represent the contribution from the spinon condensate. Consequently, the one-loop diagram (a) generates spectral weight from the two-spinon continuum, while the tree-level diagrams (b)-(c) generate  $\delta$-peaks in the DSSF  (poles of the magnetic susceptibility) at the single-spinon energy $\omega =\epsilon_{{\bm q}n}$.
 
The Green's function ${\cal G}_c$ is a rank-$1$ matrix for the simple BEC ground state given by Eq.~\eqref{green_c} with $| c_{\bm Q} \rangle$ being an $2N$-vector now. The non-condensed Green's function is given by Eq.~\eqref{green_n}, where $g_{{\bm q}n}^{+} = |n,\bm Q+\bm q \rangle \langle n,\bm Q+\bm q |$, and $g_{{\bm q}n}^{-} = P(g_{{-\bm q}n}^{+})^*P$.
The tree-level contribution to $\chi^{\mu\nu}_{\rm SP}$, i.e., the second and third terms of Eq.~\eqref{eq:chisp},
can be written in a simple form by using the following identity
\begin{eqnarray} \label{eq:trace}
\text{Tr}[F] = \sum_{n=1}^{2N} \gamma^0_{nn} \langle n \rvert \gamma^0 F \rvert n \rangle,
\end{eqnarray}
where the state vectors are normalized according to $\gamma_{nm}^0 \equiv \langle n \rvert \gamma^0 \rvert m \rangle $ and $F$ is an
arbitrary $2N \times 2N$ complex matrix. By using this relation, we obtain
 \begin{align}
\chi^{\mu\nu}_{SP;b}({\bm q},i\omega_n) & =  \frac{1}{2} \langle c_{\bm{Q}} \rvert u^{\mu} {\cal G}_n(\bm{Q}+\bm{q},i\omega_n) u^{\nu} \rvert c_{\bm{Q}} \rangle,  \\
\chi^{\mu\nu}_{SP;c}({\bm q},i\omega_n) & =  \frac{1}{2} \langle c_{\bm{Q}} \rvert u^{\nu} {\cal G}_n(\bm{Q}-\bm{q},-i\omega_n) u^{\mu} \rvert c_{\bm{Q}} \rangle,\label{eq:chisp2}
\end{align}
where $\chi^{\mu\nu}_{SP;b}({\bm q},i\omega_n)$ is the contribution from diagram (b) of Fig.~\ref{fig:chisp} 
that has poles at $i\omega_n = \varepsilon_{{\bm Q}+{\bm q} n}$, while 
$\chi^{\mu\nu}_{SP;c}({\bm q},i\omega_n)$ is the contribution from diagram (c) of Fig.~\ref{fig:chisp}
that has poles at $i\omega_n = \varepsilon_{{\bm Q}-{\bm q} n}$, $n=1,...,N$. 

Particle-hole symmetry dictates that $\rvert c_{\bm{Q}} \rangle$, the spinon Green's function and the external/internal vertices must satisfy
\begin{align}
P \left( \rvert c_{\bm Q} \rangle \right)^* & = \rvert c_{\bm Q} \rangle, & P {\cal G}_n(-{\bm q},-\omega) ^T P &= {\cal G}_n({\bm q},\omega), \label{ph1} \\
P\left( u^{\mu} \right)^T P &= u^{\mu}, & P\left( v_{\alpha} ({\bm k},{\bm q}) \right)^T P &= v_{\alpha} (-{\bm q},-{\bm k}). \label{ph2}
\end{align}
From these relations and the property ${\bm Q} \equiv -{\bm Q}$, we can demonstrate that $\chi^{\mu\nu}_{SP;b} = \chi^{\mu\nu}_{SP;c}$.
The residue associated with the SP pole {$\varepsilon_{{\bm Q}+{\bm q}n}= \varepsilon_{{\bm q}n} $ } is 
\begin{align}
Z_{\rm sp}^{\mu\nu}(n,{\bm q}) &= \lim_{\omega \rightarrow \varepsilon_{{\bm q}n}  } (\varepsilon_{{\bm q}n} - \omega) {\chi}^{\mu\nu}_{\rm SP}(\bm{q}, \omega)  = g_{\mu}^{{\bm Q}} (n,{\bm q})  \overline{g^{{\bm Q}}_{\nu}}(n,{\bm q}),
\label{residueSP}
\end{align}
where
\begin{align}
g_{\mu}^{{\bm Q}} (n,{\bm q})  &= \langle c_{\bm{Q}} \rvert u^{\mu} \rvert n,{\bm Q}+{\bm q}\rangle, \\
 \overline{g^{{\bm Q}}_{\mu}}(n,{\bm q}) &= \langle n,{\bm Q}+{\bm q} \rvert u^{\mu} \rvert c_{\bm{Q}}\rangle.
\end{align}

Diagram (d)  of Fig.~\ref{fig:chisp} contributes only to the static and uniform (in the twisted reference frame) component of the dynamical spin structure factor.
Its spectral weight is proportional to the number of lattice sites, implying that it diverges in the thermodynamic limit. 
On a finite lattice with a symmetry breaking field $\propto {\cal N}_s^{-1}$,
it is given by
\begin{eqnarray}
\chi^{\mu \nu}_{SP;d} ({\bm q},i\omega_n) &=& {1\over 2}{\cal N}_s \delta_{{\bm q},{\bm 0}} \langle c_{{\bm Q}}\rvert u^{\mu} \rvert c_{\bm Q}\rangle 
\langle c_{{\bm Q}}\rvert u^{\nu} \rvert c_{\bm Q}\rangle
\nonumber \\ 
&\times& \left( {1\over 2\epsilon_{c} - i\omega_n} + {1\over 2\epsilon_{c} + i\omega_n}  \right),
\label{chispstat}
\end{eqnarray}
where $\epsilon_c$ is of order $1/{\cal N}_s$.

\subsubsection{Square lattice Heisenberg model}

As an example, we apply the above results to the square lattice SU(2)-invariant antiferromagnetic Heisenberg  model. The external vertices are independent of frequency and momentum:
\begin{eqnarray}
u^{x}=\frac{1}{2}\left(\begin{array}{cc}
\sigma_{x} & 0\\
0 & \sigma_{x}^{T}
\end{array}\right),
\end{eqnarray}
\begin{eqnarray}
u^{y}=\frac{1}{2}\left(\begin{array}{cc}
\sigma_{y} & 0\\
0 & \sigma_{y}^{T}
\end{array}\right), 
\end{eqnarray}
\begin{eqnarray}
 u^{z}=\frac{1}{2}\left(\begin{array}{cc}
\sigma_{z} & 0\\
0 & \sigma_{z}^{T}
\end{array}\right).
\end{eqnarray}

In particular, we consider the semi-classical limit of the ground state $\rvert G_3\rangle$ by taking $S\rightarrow \infty$, for which the dynamical magnetic susceptibility
is exactly described by the linear spin wave theory (LSWT).
For $S\rightarrow \infty$, the SP approximation of the Schwinger boson theory gives $n_c=2S$, $A_x=2S$, $B_x=0$ and $\lambda = 4JS(1-\alpha)$, which makes ${\cal G}_n \sim S^{-1}$ and ${\cal G}_c \sim S^{0}$. Consequently, the DSSF has zero spectral weight in the two-spinon continuum
and finite spectral weight at the single-spinon dispersion (poles of the SP Green's function).
In other words, in the large-$S$ limit the contribution from the diagram shown in Fig.~\ref{fig:chisp} (a) is negligible in comparison to the contributions from the diagrams shown Fig.~\ref{fig:chisp} (b-d).
After performing the analytic continuation $i\omega_n \rightarrow \omega+i0^+$ to real frequency and converting the result back to the original reference frame, we obtain
\begin{align}
\chi^{xx}_{\rm SP}(\bm{q},\omega) &= \chi^{yy}_{\rm SP}(\bm{q},\omega)   
 \nonumber \\ & = S\sqrt{\frac{1-\gamma^+_{{\bm q}}}{1+\gamma^+_{{\bm q}}}} {1 \over 2} \left( {1\over \varepsilon_{\bm{q}} - \omega-i0^+} + {1\over \varepsilon_{\bm{q}} + \omega+i0^+} \right) 
, \label{eq:sxy_sp} \\
\chi^{zz}_{\rm SP}(\bm{q},\omega) & = \chi^{xx}_{\rm SP}(\bm{q},\omega) \nonumber \\ 
&  +{ {\cal N}_s n_c^2 \over 2} \delta_{{\bm q},{\bm \pi}} \left( {1\over 2\varepsilon_c - \omega-i0^+} + {1\over 2\varepsilon_c + \omega+i0^+} \right),
\label{eq:szz_sp}
\end{align}
where $\varepsilon_{\bm{q}} = (1-\alpha) \omega_{\bm q}$ and $\omega_{\bm q}$ is the single-magnon dispersion that is obtained with LSWT for the square lattice AFM Heisenberg model. 
We note that the diagrams shown  in Fig.~\ref{fig:chisp} (b) and (c) produce an SU(2) invariant contribution to the magnetic susceptibility that is the same for the three different  ground states, while the diagram shown in Fig.~\ref{fig:chisp} (d) breaks the SU(2) invariance explicitly and leads to the finite difference $\chi^{zz}_{\rm SP}(\bm{q},\omega) - \chi^{xx/yy}_{\rm SP}(\bm{q},\omega)$ for 
the ground states $\rvert G_2\rangle$ and $\rvert G_3\rangle$.
We also note that
different choices of $\alpha$, i.e., different decoupling schemes of the original Hamiltonian, lead to different SP approximations:
$\chi^{xx}_{\rm SP}(\bm{q},\omega)$ and $\chi^{yy}_{\rm SP}(\bm{q},\omega)$ coincide with the LSWT result  only for $\alpha=0$ \cite{auerbach1988spin}.


There is also an important qualitative difference  between the SP approximation of the SB theory and LSWT.
 The inelastic contribution to the SP susceptibility, arising  from the bubble diagrams {shown in Fig.~\ref{fig:chisp}~(b) and (c)}, is 
 SU(2) invariant. This property leads to the result given in Eqs.~(\ref{eq:sxy_sp}) and (\ref{eq:szz_sp}), 
implying that the dynamical spin structure factor satisfies
 $S^{xx}({\bm q},\omega\neq 0) = S^{yy}({\bm q},\omega\neq 0) =S^{zz}({\bm q},\omega \neq 0)$.
This \emph{qualitatively incorrect} result must be contrasted with the LSWT theory, where only the transverse spin susceptibility has
poles corresponding to magnons, while the longitudinal susceptibility only  includes a two-magnon continuum with relative small spectral weight.
This simple observation  invalidates the association of the poles of the SP susceptibility with magnon modes.
As we will show in Sec.~\ref{MP}, the correct LSWT result can be recovered by including corrections due to fluctuations around the SP solution~\cite{zhang2019large}.

\section{Beyond the saddle-point approximation}
\label{BSP}

Fluctuations of the auxiliary fields ($W,{\bar W}$) and the Lagrangian multiplier ($\lambda$) around the SP configuration are described by the  fluctuation fields
\begin{gather}
\phi_{\alpha} (q) = \Big [ W_{{\bm \delta}}^{\mu}(q) -W_{{\bm \delta}}^{\mu}(q)\rvert_{\rm SP}, {\bar W}_{{\bm \delta}}^{\mu}(-q) - {\bar W}_{{\bm \delta}}^{\mu}(-q)\rvert_{\rm SP},\nonumber  \\
\lambda (q) -\lambda(q)\rvert_{\rm SP} \Big ],
\end{gather}
where $\alpha$ represents the field indices ($W_{{\bm \delta}}^{\mu}$, ${\bar W}_{{\bm \delta}}^{\mu}$ or $\lambda$ ).

The large\ssz{-}$N$ expansion is then obtained by expanding the effective action Eq.~(\ref{action2}) around the $\rm SP$ in powers of the fields $\phi_{\alpha} (q)$:
\begin{gather}
S_{\rm eff} = S_{\rm eff}^{\rm SP} + {1 \over 2} \sum_{q,\alpha_1,\alpha_2}S^{(2)}_{\alpha_1,\alpha_2}(q) \  \phi_{\alpha_1}(-q)\phi_{\alpha_2}(q) + S_{\rm int} ,
\end{gather}
with
\begin{gather}
S_{\rm int} = \sum_{n \geq 3}^{\infty} {1\over n!} \sum_{\substack{ {\alpha_1...\alpha_n} \\ {q_1...q_n}} } 
S_{\alpha_1...\alpha_n}^{[n]}(q_1,...,q_n) \ \phi_{\alpha_1}(q_1)...\phi_{\alpha_n}(q_n), \nonumber
\end{gather}
where the first term is the SP contribution, the second (quadratic) 
term corresponds to the usual random phase approximation (RPA), and the last term describes the interaction between the fluctuation fields \cite{zhang2019large,book2012interacting}.

The coefficients of the quadratic form can be expressed as
\begin{equation}
    S^{(2)}(q)=\Pi_0 - \Pi(q) ,
\end{equation}
where
\begin{eqnarray}
(\Pi_0)_{\alpha_1 \alpha_2} &=& {1-\alpha \over 4J_{\delta}} (\delta_{\alpha_1,\bar{W}_{\delta}^{A}} \delta_{\alpha_2,{W}_{\delta}^{A}} + \delta_{\alpha_1,{W}_{\delta}^{A}} \delta_{\alpha_2,\bar{W}_{\delta}^{A}} ) \nonumber \\
&& + {\alpha \over 4J_{\delta}} (\delta_{\alpha_1,\bar{W}_{\delta}^{B}} \delta_{\alpha_2,{W}_{\delta}^{B}} + \delta_{\alpha_1,{W}_{\delta}^{B}} \delta_{\alpha_2,\bar{W}_{\delta}^{B}} ) ,
\end{eqnarray}
and the polarization operator is 
\begin{gather}
\Pi_{\alpha_1\alpha_2} (q)={1\over2{\cal N}_s\beta} \sum_{k}{1\over N} \text{Tr}[{\cal G}_{ \rm SP}(k) v_{\alpha_1}({\bm k}, {\bm k}+{\bm q}) {\cal G}_{ \rm SP}(k+q) \nonumber \\ 
v_{\alpha_2}({\bm k}+{\bm q},{\bm k})].\label{polarization1} 
\end{gather}
The RPA propagator of the fluctuation fields is given by
\begin{align}
{1\over N} [D_{\alpha_1\alpha_2}] (q) &= {1 \over {\cal Z}^{(2)}} \int {\cal D}[\phi] \phi_{\alpha_1}(q) \phi_{\alpha_2}(-q) \nonumber \\
& \times e^{ -{N \over 2} \sum\limits_{q,\alpha_1,\alpha_2} S^{(2)}_{\alpha_1,\alpha_2}(q) \  \phi_{\alpha_1}(-q)\phi_{\alpha_2}(q) } \nonumber \\
&=[(N S^{(2)})^{-1}]_{\alpha_1\alpha_2},
\end{align}
where 
\begin{equation}
{\cal Z}^{(2)} = \int {\cal D}[\phi]  e^{-{N \over 2} \sum\limits_{q,\alpha_1,\alpha_2} S^{(2)}_{\alpha_1,\alpha_2}(q) \  \phi_{\alpha_1}(-q)\phi_{\alpha_2}(q) }. 
\end{equation}

 
The coefficients of the higher-order corrections ($n\geq 3$)
are 
\begin{gather}
S_{\alpha_1...\alpha_n}^{[n]}(q_1,...,q_n)={(-1)^{n+1} \over n ({\cal N}_s\beta)^{n\over2} } {1\over N} \sum_{P} 
\text{Tr}[{\cal G}_{\rm SP} v_{P_1}... {\cal G}_{\rm SP} v_{P_n}],
\end{gather}
where $\sum_{P}$ runs over all permutations of the particle index $(\alpha_i,q_i), 1\leq i \leq n$, which symmetrizes the vertex function relative to the exchange of any pair of particles.

\subsection{RPA propagator}

\begin{figure}[!t]
\centering
\includegraphics[scale=0.18]{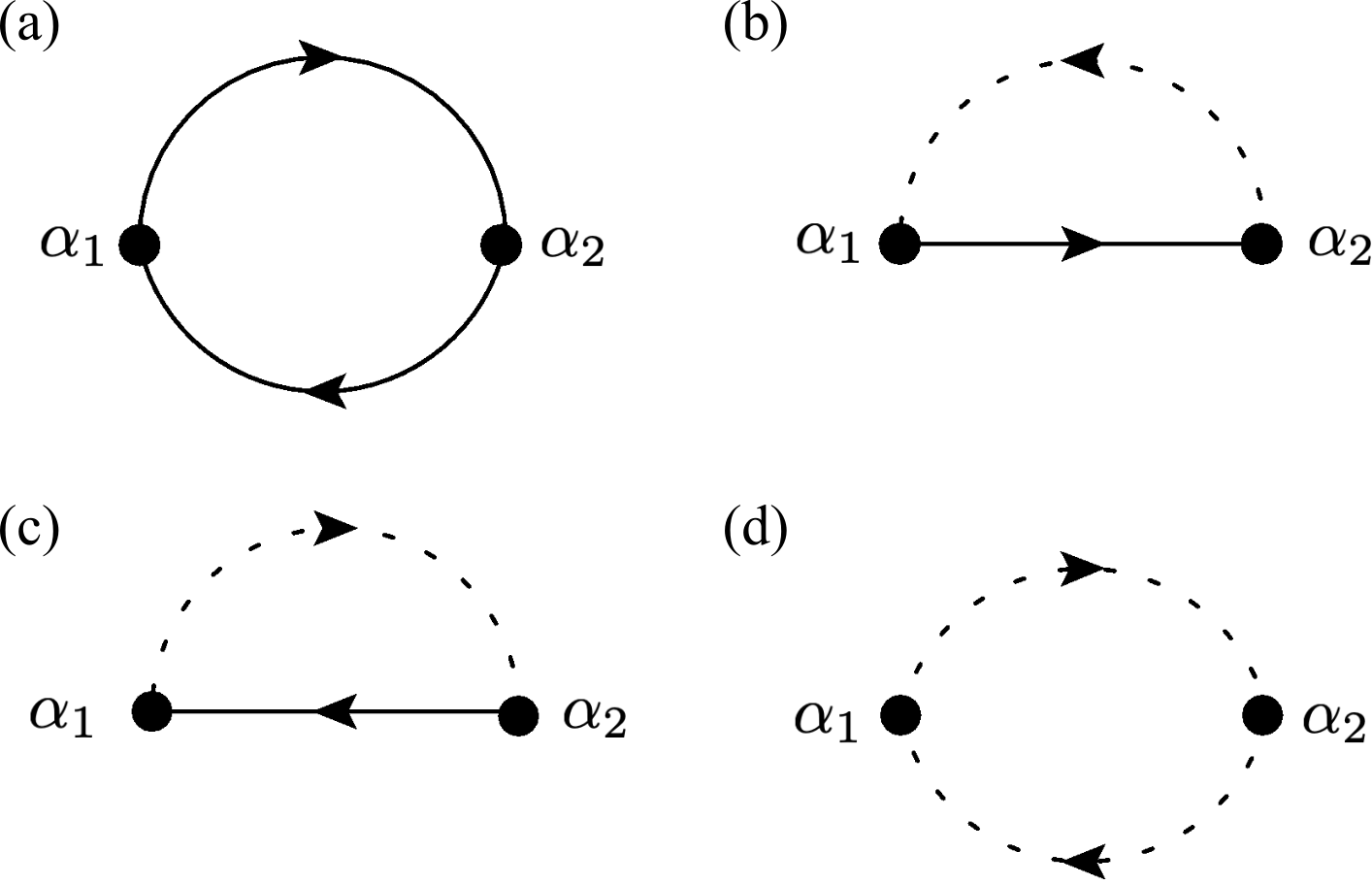}
\caption{Diagramatic representation of the four contributions to the  polarization operator $\Pi({\bm q},i\omega_n)$
corresponding to the four terms of Eq.~\eqref{polarization}. The full circles represent the internal vertices of the theory introduced in Eq.~\eqref{intvert}.}
\label{fig:polarization}
\end{figure}

Because of the spinon condensation, the polarization operator in Eq.~(\ref{polarization1}) also includes four contributions
\begin{align}\label{polarization}
&\Pi_{\alpha_1\alpha_2} ({\bm q},i\omega_n) \nonumber \\
=&{1\over 2{\cal N}_s\beta} \sum_{k}{1\over N} \text{Tr}[{\cal G}_n(k) v_{\alpha_1}({\bm k}, {\bm k}+{\bm q}) {\cal G}_n(k+q)
v_{\alpha_2}({\bm k}+{\bm q},{\bm k})]  \nonumber \\
+&{1\over 2{\cal N}_s\beta} \sum_{k}{1\over N} \text{Tr}[{\cal G}_c(k) v_{\alpha_1}({\bm k}, {\bm k}+{\bm q}) {\cal G}_n(k+q)
v_{\alpha_2}({\bm k}+{\bm q},{\bm k})]  \nonumber \\
+&{1\over 2{\cal N}_s\beta} \sum_{k}{1\over N} \text{Tr}[{\cal G}_n(k) v_{\alpha_1}({\bm k}, {\bm k}+{\bm q}) {\cal G}_c(k+q)
v_{\alpha_2}({\bm k}+{\bm q},{\bm k})]  \nonumber \\
+&{1\over 2{\cal N}_s\beta} \sum_{k}{1\over N} \text{Tr}[{\cal G}_c(k) v_{\alpha_1}({\bm k}, {\bm k}+{\bm q}) {\cal G}_c(k+q)
v_{\alpha_2}({\bm k}+{\bm q},{\bm k})]. 
\end{align}
The first term corresponds to the one-loop Feynman diagram shown in Fig.~\ref{fig:polarization} (a). The rest of the terms include at least one contribution from the condensed spinons, implying  that they only exist in the magnetically ordered phase.
In particular, the second and the third terms, represented by the tree-level diagrams shown in Figs.~\ref{fig:polarization} (b) and (c), are 
equal because of the same symmetry arguments that lead to the equivalence between the diagrams
shown in Figs.~\ref{fig:chisp} (b) and (c):
\begin{align}
&\Pi_{\alpha_1\alpha_2}^{(b)} ({\bm q},i\omega_n) = \Pi_{\alpha_1\alpha_2}^{(c)} ({\bm q},i\omega_n) \nonumber \\
=& {1\over 2N} \langle c_{{\bm Q}} \rvert v_{\alpha_1}({\bm Q}, {\bm Q}+{\bm q}) {\cal G}_n({\bm Q}+{\bm q},i\omega_n) v_{\alpha_2}({\bm Q}+{\bm q},{\bm Q}) \rvert c_{\bm Q}\rangle.   \label{eq:pib}
\end{align}
The last term of Eq.~(\ref{polarization}),  corresponding  to the diagram shown in Fig.~\ref{fig:polarization} (d),  has only zero-frequency and momentum $(\omega=0, {\bm q}={\bm 0})$ components:
\begin{multline}
\Pi_{\alpha_1\alpha_2}^{(d)} ({\bm q},i\omega_n) 
=(2\pi)^3 
\delta(\omega_n)
\delta({\bm q} ) \\
{1\over 2N} \langle c_{{\bm Q}} \rvert v_{\alpha_1}({\bm Q},{\bm Q}) \rvert c_{{\bm Q}} \rangle \langle c_{{\bm Q}} \rvert v_{\alpha_2}({\bm Q},{\bm Q}) \rvert c_{{\bm Q}}\rangle \label{eq:pid}
\end{multline}

\section{Cancellation of  single-spinon poles}
\label{CSSP}

\begin{figure*}[!t]
\centering
\includegraphics[scale=0.14]{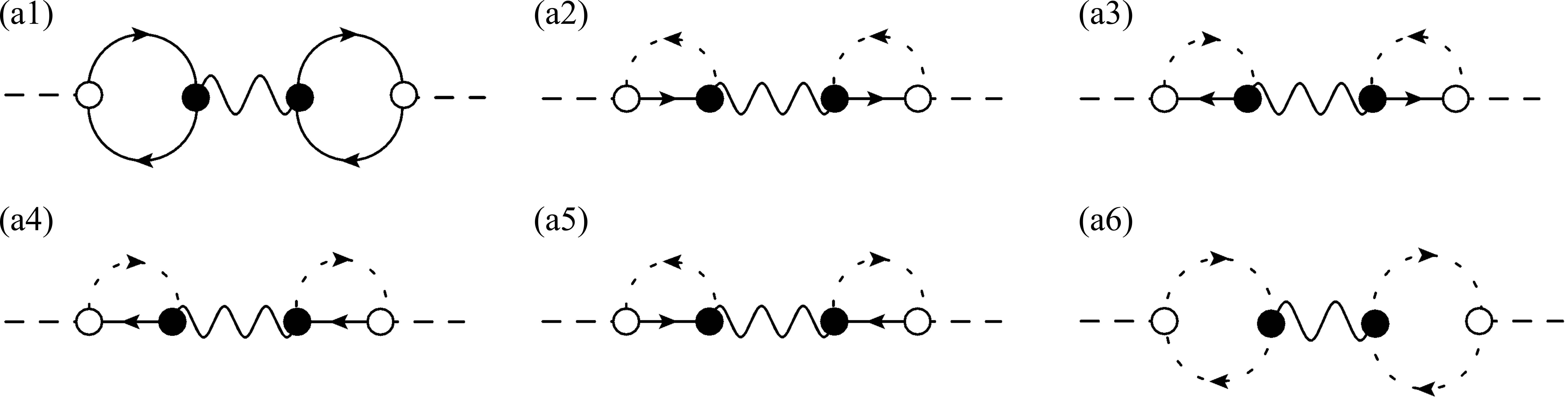}
\caption{$1/N$ diagrams of dynamical spin susceptibility. Wavy line refers to the RPA propagator ${1\over N} D({\bm q},\omega)$. }
\label{fig:chi1overN}
\end{figure*}

The primary purpose of this section is to demonstrate that the 
spectral weight of $\chi^{\mu \nu}_{SP}$ at the single-spinon poles, $\omega=\varepsilon_{\bm{q}n}$,
is exactly canceled out by a counter-diagram that is \emph{nominally} of order $1/N$.  By ``nominally'' here we mean the well-defined order $1/N^{P-L}$ ($L$ is the number of internal loops and $P$ is the number of RPA propagators) that the diagram would have \emph{in absence of a condensate}~\cite{book2012interacting}. Naively, this cancellation is unexpected because a $1/N$ contribution cannot cancel a contribution of order
$1/N^0$ for arbitrary values of $N$. The crucial observation is that the  classification of the Feynman diagrams in powers of 
$1/N$~\cite{book2012interacting} does not take into consideration singular contributions that are generated by the finite condensate fraction.
In other words, diagrams that are nominally of order $1/N$ include a singular  contribution of order ${\cal O}(1/N^0)$  whenever the condensate fraction $n_c$ is finite. This contribution corresponds to isolated poles with residues of ${\cal O}(1/N^0)$  that exactly cancel out the poles of the SP solution. 

In previous works on the triangular Heisenberg antiferromagnet, we reported this cancellation for one specific diagram of order $1/N$~\cite{zhang2019large}. 
Here we will reveal the origin of the cancellation and generalize the result to other diagrams.
From a physical point of view, this exact cancellation simply means that spinons are not  quasi-particles of the magnetically ordered state. 
The true quasi-particles are magnons (two-spinon bound states) arising from poles of the  RPA propagator~\cite{ghioldi2018dynamical,zhang2019large}.

\subsection{$1/N$ corrections}

The $1/N$ correction we considered in previous works \cite{ghioldi2018dynamical,zhang2019large} is given by
\begin{equation}
    \chi_{FL}^{\mu \nu}(q) = \sum_{\alpha_1 \alpha_2} \Lambda^{\mu \alpha_1}(q) \frac{1}{N} D_{\alpha_1 \alpha_2}(q) \Lambda^{\nu \alpha_2}(-q) ,
\end{equation}
where 
\begin{equation}
    \Lambda^{\mu \alpha_1}(q) = {1\over 2{\cal N}_s\beta} \sum_{k} \text{Tr}[{\cal G}_{\rm SP}(k) u^{\mu} {\cal G}_{\rm SP}(k+q)
v_{\alpha_1}({\bm k}+{\bm q},{\bm k})] .
\end{equation}
This contribution is represented by the diagrams shown in Fig.~\ref{fig:chi1overN}.
We will first demonstrate that the poles of $\chi^{\mu \nu}_{SP}$ are cancelled by these diagrams, which represent different contributions to a diagram of nominal order
$1/N$  in the large-$N$ expansion of the magnetic susceptibility~\cite{ghioldi2018dynamical,zhang2019large}.
We have split this diagram into a sum of different sub-diagrams because the spinon and the RPA propagators
include condensed and non-condensed contributions represented with different types of lines.
Similarly to Fig.~\ref{fig:chisp}, the full (dashed) line represents the propagator of the non-condensed (condensed)  spinons, while the black wavy line represents the RPA propagator.

Since the diagram includes two loops and each loop includes two spinon propagators, there are four tree-level diagrams
shown in Figs.~\ref{fig:chi1overN} (a2-a5) that have poles of  the non-condensed spinon line and of  the RPA propagator. 
They are  called ``tree-level'' diagrams simply because the condensed spinon line carries fixed momentum ${\bm Q}$ and 
zero Matsubara frequency $i \omega_n = 0$, which is equivalent to a classical external potential.
The  diagram shown in Fig.~\ref{fig:chi1overN}~(a1) only includes spectral weight from the two-spinon continuum, while the diagram shown in  Fig.~\ref{fig:chi1overN}~(a6) corresponds to an elastic ($\omega=0$) contribution.
Our main focus is on  the pole singularities that arise from the non-condensed single-spinon propagators in the diagrams
shown in  Figs.~\ref{fig:chi1overN} (a2-a5). The particle-hole symmetry that leads to Eqs.~(\ref{ph1}) and (\ref{ph2}),
implies that these four diagrams give identical contributions $\chi_{FL;(a2-a5)}^{\mu \nu} ({\bm q},\omega)$ 
to  the dynamical spin susceptibility with
\begin{widetext}
\begin{align}
  \Lambda_{(a2-a5)}^{\mu \alpha_1}(\bm q, \omega)  =  
   {1\over 2{\cal N}_s\beta} \sum_{k} \text{Tr}[{\cal G}_c(k) u^{\mu} {\cal G}_n(k+q) v_{\alpha_1}({\bm k}+{\bm q},{\bm k})] 
   + {1\over 2{\cal N}_s\beta} \sum_{k} \text{Tr}[{\cal G}_n(k) u^{\mu} {\cal G}_c(k+q) v_{\alpha_1}({\bm k}+{\bm q},{\bm k})]. \nonumber 
\end{align}
Consequently,
\begin{align}
  \chi_{FL;(a2-a5)}^{\mu \nu} ({\bm q},\omega) =  {1\over 4 N} & \left[ \langle c_{{\bm Q}} \rvert u^{\mu} {\cal G}_n({\bm q}+{\bm Q},\omega) v_{\alpha_1} ({\bm q}+{\bm Q},{\bm Q}) 
\rvert c_{{\bm Q}} \rangle +  \langle c_Q\rvert  v_{\alpha_1}({\bm Q},-{\bm q}+{\bm Q}) {\cal G}_n(-{\bm q}+{\bm Q},-\omega) u^{\mu}  \rvert c_{{\bm Q}} \rangle \right]  \nonumber \\
 \times D_{\alpha_1\alpha_2} ({\bm q},\omega) & 
\left[  \langle c_{{\bm Q}} \rvert v_{\alpha_2}({\bm Q},{\bm q}+{\bm Q}) {\cal G}_n({\bm q}+{\bm Q},\omega)  u^{\nu} \rvert c_{{\bm Q}} \rangle
 +  \langle c_{{\bm Q}} \rvert u^{\nu} {\cal G}_n(-{\bm q}+{\bm Q},-\omega)  v_{\alpha_2}(-{\bm q}+{\bm Q},{\bm Q})  \rvert c_{{\bm Q}} \rangle \right].
\end{align}
It is clear that, like $\chi^{\mu \nu}_{\rm SP} ({\bm q}, \omega)$ (see Fig.~\ref{fig:chisp}), $\chi_{FL;(a2-a5)}^{\mu \nu} ({\bm q},\omega)$ has poles at the single-spinon energies {$\epsilon_{\bm{q} n}$} arising from loops with one dashed and one full line.
For the four diagrams, the residue of this pole  reads
\begin{align} \label{eq:Z1overN}
Z_{FL}^{\mu\nu} (n,{\bm{q}}) &= \lim_{\omega \rightarrow {\epsilon_{\bm{q} n}} } (\varepsilon_{{\bm q}n} - \omega) {\chi}^{\mu\nu}_{FL}(\bm{q}, \omega) 
= \frac{1}{- i \eta N}
g_{\mu}^{{\bm Q}} (n,{\bm q})  \overline{g^{{\bm Q}}_{\nu}}(n,{\bm q})
\langle n,\bm{Q}+\bm{q} \rvert \Sigma(\bm{Q}+\bm{q}, { \epsilon_{\bm{q} n}})  \rvert n,\bm{Q}+\bm{q}\rangle,
\end{align}
where $\eta$ is an infinitesimal positive number which arises from the pole of the single-spinon Green's function, $\lim_{\omega \rightarrow {\epsilon_{\bm{q} n}}} 1/({\epsilon_{\bm{q} n}} - \omega - i\eta)$,
 and the matrix element is given by
\begin{eqnarray}
\langle n,\bm{Q}+\bm{q} \rvert  \Sigma({\bm Q}+\bm{q},{ \epsilon_{\bm{q} n}})   \rvert n,\bm{Q}+\bm{q}\rangle
& = & \sum_{\alpha_1 \alpha_2} 
D_{\alpha_1 \alpha_2}(\bm{q}, {\epsilon_{\bm{q} n}}) {\bar f}_{\alpha_1}^{{\bm Q}}(n,{\bm q}) f^{{\bm Q}}_{\alpha_2}(n,{\bm q})=
 {\bar {\bm f}}^{{\bm Q}}(n,{\bm q})  D(\bm{q}, { \epsilon_{\bm{q} n}}) {\bm f}^{{\bm Q}}(n,{\bm q}),
 \label{eq:matel}
\end{eqnarray}
where we have introduced the row and column vectors:
\begin{equation}
{\bm f}^{\bm Q} (n,{\bm q}) =
[ \langle c_{{\bm Q}} \rvert v_{1}({\bm Q}, {\bm Q}+{\bm q}) \rvert n,\bm{Q} + \bm{q}\rangle, ... ,
\langle c_{{\bm Q}} \rvert v_{ N_{\phi} } ({\bm Q}, {\bm Q}+{\bm q}) \rvert n,\bm{Q} + \bm{q}\rangle],
\;\;\;
\bar{{\bm f}}^{{\bm Q}}(n,{\bm q}) =
\left [
\begin{array}{ccc}
 \langle n,\bm{Q}+\bm{q}\rvert v_{1}({\bm Q}+{\bm q},{\bm Q}) \rvert c_{\bm Q}\rangle \\
 . \\
 . \\
 . \\
 \langle n,\bm{Q}+\bm{q}\rvert v_{N_{\phi}}({\bm Q}+{\bm q},{\bm Q}) \rvert c_{\bm Q}\rangle
 \end{array}
 \right ]
\end{equation}
where $N_{\phi}$ is the number of flavors of the fluctuation fields $\phi_{\alpha}$.
We note that ${\bm f}^{\bm Q} (n,{\bm q})$ and $\bar{{\bm f}}^{{\bm Q}}(n,{\bm q})$ are both proportional to the factor $\sqrt{n_c}$ carried by the ket $\rvert c_{\bm Q}\rangle$.

 According to Eq.~\eqref{residueSP}, the condition for the cancellation of the pole of  $\chi^{\mu \nu}_{SP} ({\bm q}, \omega)$ at ${\epsilon_{\bm{q} n}}$ is that the matrix element $\langle n,\bm{Q}+\bm{q} \rvert \Sigma(\bm{Q}+\bm{q}, {\epsilon_{\bm{q} n}})  \rvert n,\bm{Q}+\bm{q}\rangle$ must be equal to $i\eta N$.
In other words, as we will demonstrate below, ${ \epsilon_{\bm{q} n}}$ must be a \textit{zero} of the propagator $D(\bm{q}, \omega)$.
\end{widetext}

\subsubsection{Anomalous large-$N$ scaling of RPA propagator}

The RPA propagator is given by the inverse of the fluctuation matrix:
\begin{equation}
D(\bm{q},\omega)= \left[ \Pi_{0}-\Pi(\bm{q},\omega) \right]^{-1}.
\end{equation}
The tree-level diagrams of the polarization operator $\Pi(\bm{q}, \omega)$ (see Fig.~\ref{fig:polarization}) have poles at
$\omega={\epsilon_{\bm{q} n}}$,
namely
\begin{equation}
\Pi_{\alpha_1\alpha_2}^{(b)} ({\bm q}, {\epsilon_{\bm{q} n}} )
= \Pi_{\alpha_1\alpha_2}^{(c)} ({\bm q}, {\epsilon_{\bm{q} n}} )
= {1\over 2N} {1\over (-i\eta)} f_{\alpha_1}^{{\bm Q}} (n,{\bm q}) \bar{f}_{\alpha_2}^{{\bm Q}} (n,{\bm q}). \label{eq:pib2}
\end{equation}
It is important to note that only the spinon from the  $n$-band with momentum ${\bm Q}+{\bm q}$ contributes to
the polarization operator, implying that  $\Pi(\bm{q},{\epsilon_{\bm{q} n}})\sim{\cal O}(1/N)$. 

\begin{figure}[!t]
\centering
\includegraphics[scale=0.25]{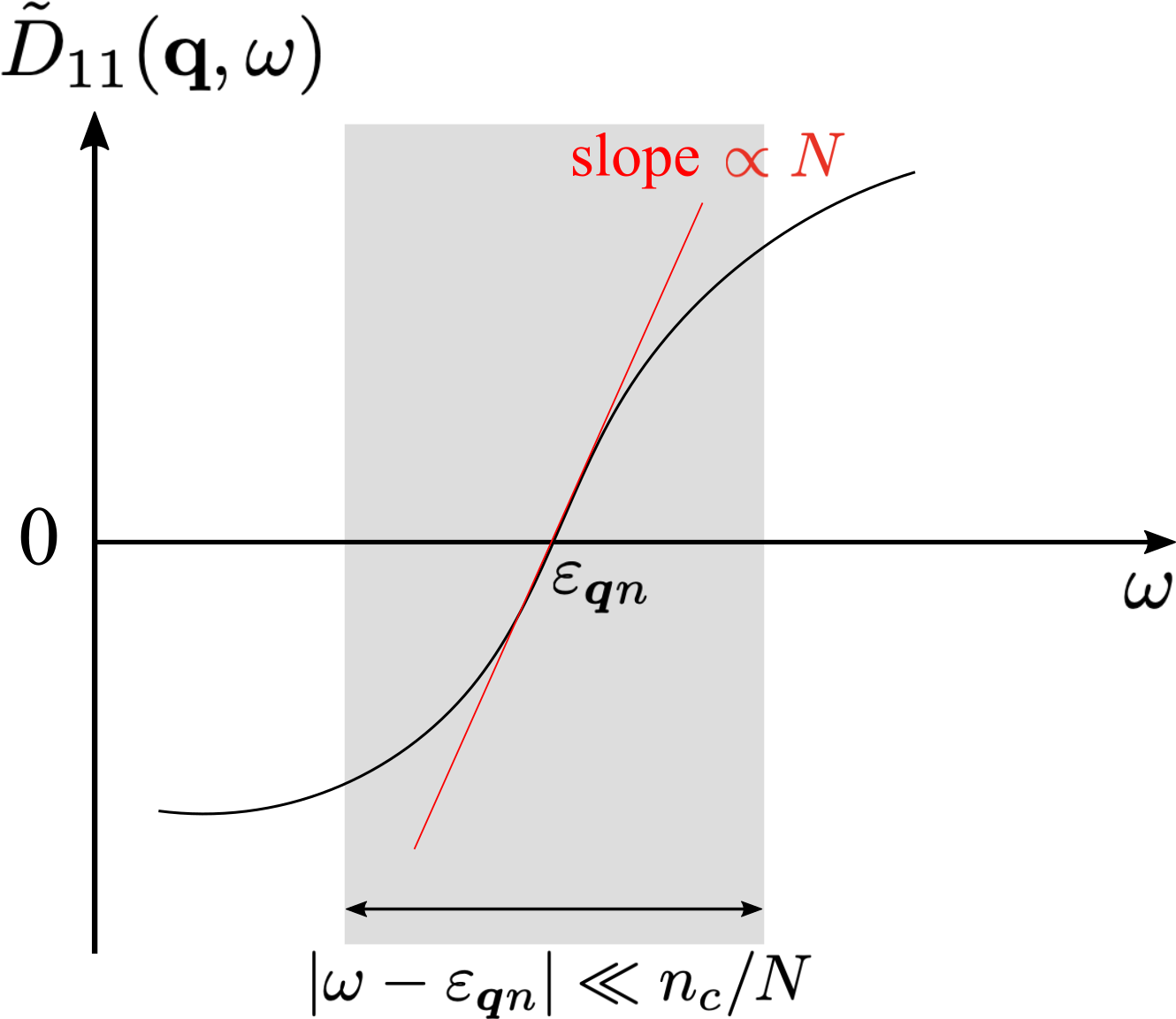}
\caption{Vicinity of the zero of $\tilde{D}_{11}({\bm q},\omega)$ (gray area) where the RPA propagator displays anomalous large-$N$ scaling, namely, $d \tilde{D}_{11} / d \omega \rvert_{\omega=\varepsilon_{{\bm q} n}} \propto {N}$, as indicated by the red line.}
\label{fig:Drpa_anormalousN}
\end{figure}

Since the tree-level diagrams diverge at $\omega={\epsilon_{\bm{q} n}}$, we can neglect the regular contribution from the loop diagram shown in Fig.~\ref{fig:polarization}~{(a)} and the polarization operator is dominated by a rank-$1$ matrix
\begin{eqnarray}\label{eq:pi_rank1}
\Pi_{\alpha_1\alpha_2} (\bm{q},{\epsilon_{\bm{q} n}})  = 
\frac{ f_{\alpha_1}^{{\bm Q}} (n,\bm{q}) \bar{f}_{\alpha_2}^{{\bm Q}}(n,\bm{q})}{-i \eta N}.
\end{eqnarray}
To compute the RPA propagator, it is convenient to introduce  new pair of orthogonal bases 
$\{ {\boldsymbol \mu}_{\alpha} \}$ and $\{ {\boldsymbol \nu}^{\dagger}_{\alpha} \}$ such that the first vector of each basis  are unit vectors 
parallel to ${\bm f}^{\bm Q} (n,{\bm q})$ and $\bar{{\bm f}}^{{\bm Q} }(n,\bm{q})$, respectively:
\begin{eqnarray}
 {\boldsymbol \mu}_{1} =\frac{{\bm f}^{{\bm Q}}(n,\bm{q})}{\left\Vert {\bm f}^{{\bm Q}}(n,\bm{q})\right\Vert }, \;\;\;\;
 {\boldsymbol \nu}_{1}^{\dagger}=\frac{\bar{{\bm f}}^{{\bm Q}}(n,\bm{q})}{\left\Vert \bar{{\bm f}}^{{\bm Q}}(n,\bm{q})\right\Vert },
\end{eqnarray}
where $\Vert {\bm f} \Vert$ is the norm of the vector ${\bm f}$. The basis $\{ {\boldsymbol \mu}_{\alpha} \}$ spans the left space of the fluctuation matrix $\Pi_0 - \Pi$, while
$\{ {\boldsymbol \nu}^{\dagger}_{\alpha} \}$ spans the right space.
In the new basis, the fluctuation matrix is given by
\begin{eqnarray}
\!\!\!\!\!\! \tilde{D}^{-1}(\bm{q},{\epsilon_{\bm{q} n}} ) &=& U^{\dagger}D^{-1}(\bm{q}, \epsilon_{\bm{q} n} ) V
\nonumber \\
& =  &
\frac{\left\Vert {\bm f}^{{\bm Q}}(n,\bm{q})\right\Vert \left\Vert 
\bar{\bm f}^{{\bm Q}}(n,\bm{q})\right\Vert }{ i\eta N}  I^{(1,1)}
+{\cal O}(\eta^0),
\label{eq:inverseD_rotated}
\end{eqnarray}
where $I^{(1,1)}$ is the single-entry $N_{\phi} \times N_{\phi}$ matrix 
$I^{(1,1)}_{\alpha_1, \alpha_2} = \delta_{\alpha_1, 1} \delta_{\alpha_2, 1}$
and the column $j$ of the unitary matrix $U$ ($V$) is equal to the
vector ${\bm u}_{j}$ (${\bm v}_{j}$):
\begin{eqnarray}
U  = \left [{\boldsymbol \mu}_{1},..., {\boldsymbol \mu}_{N_{\phi}}  \right ],\; \;\;\;
V =  \left [{\boldsymbol \nu}_{1},..., {\boldsymbol \nu}_{N_{\phi}} \right ].
\end{eqnarray}
The ($1,1$) component $[\tilde{D}^{-1}(\bm{q},{\epsilon_{\bm{q} n}} )]_{11}$ scales as ${n_c / N}$ instead of the usual scaling $\sim N^0$ for a magnetic disordered state ($n_c=0$). Since $n_c/(\eta N) \gg 1$ for an ordered magnet with $n_c \neq 0$,  the inverse of Eq.~\eqref{eq:inverseD_rotated} leads to
\begin{eqnarray}
\!\!\!\!\!\!  \tilde{D}_{11}({\bm q},{\epsilon_{\bm{q} n}}) &\equiv&  {{\boldsymbol  \nu}}^{\dagger}_1  D({\bm q},{\epsilon_{\bm{q} n}}) 
{{\boldsymbol  \mu}}_1  \nonumber \\
& = & \frac{i\eta N}{ \left\Vert {\bm f}^{{\bm Q}}(n,\bm{q})\right\Vert \left\Vert \bar{\bm f}^{{\bm Q}}(n,\bm{q})\right\Vert }
+{\cal O}({\eta^{2}N^2 \over n_c^{2}}),
\label{expD}
\end{eqnarray}
implying that $\tilde{D}_{11}({\bm q},{\epsilon_{\bm{q} n}}) \propto N$ instead of the  $N^0$ scaling that is obtained in the absence of spinon condensation. Since $i\eta = \omega-\varepsilon_{\bm q n}$ is an infinitesimal number, $\omega=\varepsilon_{{\bm q}n}$ is a \textit{zero} of $\tilde{D}_{11}({\bm q}, \omega )$ and the ${\cal O}(N)$ scaling holds in the vicinity of this zero, $|\omega-{\epsilon_{\bm{q} n}}| \ll n_c/N$, where the polarization operator is dominated by the contribution from  the $n$-th band spinon [see Eq.~(\ref{eq:pi_rank1})]. 
As a result,  $d\tilde{D}_{11}({\bm q}, {\omega })/ d\omega \rvert_{\omega=\varepsilon_{{\bm q}n}} \propto N$
as illustrated in Fig.~\ref{fig:Drpa_anormalousN}. 
For $|\omega-{\epsilon_{\bm{q} n}}| \gg n_c/N$, $D({\bm q},\omega)$ recovers the usual scaling $D({\bm q},\omega) \propto N^0$.
A similar analysis shows that 
$\tilde{D}_{1,\sigma^{\prime}>1}$ and $\tilde{D}_{ \sigma>1,1}$ have the same anomalous large-$N$ scaling in the vicinity of $\omega={\epsilon_{\bm{q} n}}$, while the other components $\tilde{D}_{ \sigma>1,\sigma^{\prime}>1}$ satisfy the usual scaling $\tilde{D} \propto  N^0$.

\subsubsection{Cancellation of single-spinon poles}

By using the result in Eq.~(\ref{expD}), the bilinear form Eq.~\eqref{eq:matel} can be expressed as
\begin{eqnarray}
   \left\Vert \bar{\bm f}^{{\bm Q}}(n,{\bm q}) \right\Vert  \left\Vert {\bm f}^{{\bm Q} } (n, {\bm q}) \right\Vert
\tilde{D}_{11}(\bm{q},{\epsilon_{\bm{q} n}}) +{\cal O}(\eta^{2}),
\end{eqnarray}
and 
\begin{eqnarray}
\!\!\! Z_{FL}^{\mu\nu} (n,{\bm{q}}) &=& \frac{1}{- i \eta N}
g_{\mu}^{{\bm Q}} (n,{\bm q})  \overline{g^{{\bm Q}}_{\nu}}(n,{\bm q}) 
\nonumber \\
&& \left\Vert \bar{\bm f}^{{\bm Q}}(n,{\bm q}) \right\Vert  \left\Vert {\bm f}^{{\bm Q} } (n, {\bm q}) \right\Vert   \tilde{D}_{11}(\bm{q},{\epsilon_{\bm{q} n}}).
\label{resi}
\end{eqnarray}
The key observation is that the residue $Z_{FL}^{\mu\nu} (n,{\bm{q}})$ is of order $1/N^0$ although it represents a (singular) contribution to a $1/N$ correction of the dynamical spin susceptibility that is diagramatically represented in
Fig.~\ref{fig:chi1overN} . We note that this anomaly arises from the 
first term of Eq.~\eqref{expD}: the propagator $D({\bm q}, \omega)$, which is nominally of order $1/N^0$, has a contribution of order $N$ that is proportional to $\eta$. According to Eqs.~\eqref{expD} and \eqref{resi}, we have
\begin{align}
Z_{FL}^{\mu\nu} (n,{\bm{q}}) = 
- g_{\mu}^{{\bm Q}} (n,{\bm q})  \overline{g^{{\bm Q}}_{\nu}}(n,{\bm q}).
\end{align}
In other words, the  anomalous scaling of $Z_{FL}^{\mu\nu} (n,{\bm{q}})$ leads to an exact cancellation with the residue of $\chi^{\mu \nu}_{SP} ({\bm q}, \omega)$ at $\omega=\epsilon_{\bm{q} n}$ that is given in Eq.~\eqref{residueSP}:
\begin{align}
Z_{FL}^{\mu\nu}(n,{\bm q}) = - Z_{\rm sp} ^{\mu\nu}(n,{\bm q}),
\label{cancellation}
\end{align}
or
\begin{align}
    \lim_{\omega \rightarrow \epsilon_{{\bm q} n}} (\epsilon_{{\bm q} n} - \omega) \chi_{FL}^{\mu \nu}(\bm q, \omega) = -  \lim_{\omega \rightarrow \epsilon_{{\bm q} n}} (\epsilon_{{\bm q} n} - \omega) \chi_{\rm SP}^{\mu \nu}(\bm q, \omega).
\end{align}
Eq.~\eqref{cancellation} is one of the key results of this work, which has a clear physical meaning:
single spinon excitations are not true quasi-particles of the magnetically ordered state.
The true collective modes are magnons, which arise from  the poles of the RPA propagator $D({\bm q}, \omega)$.
As we will discuss below, this important result implies that adding \emph{all the diagrams} up to a given \emph{nominal order} in $1/N$ is not the correct strategy to obtain qualitatively correct results in the presence of a condensate. For each diagram of a given order in $1/N$, there is  a ``counter-diagram'' of higher nominal order, which must be added  to cancel the unphysical single-spinon poles.

Finally,  the diagram  shown in Fig.~\ref{fig:chi1overN} (a6) contributes to the static and uniform magnetic susceptibility:
\begin{multline}
\chi_{FL;(a6)}^{\mu \nu}({\bm q},\omega) =  \chi_{SP;d}^{\mu \nu}({\bm q},\omega) \\
\times \sum_{\alpha_1\alpha_2} D_{\alpha_1\alpha_2} ({\bm 0},\omega)  f_{\alpha_1}^{ {\bm Q} }(n,\bm 0) f_{\alpha_2}^{ {\bm Q} }(n,\bm 0) \\
\times {{\cal N}_s \over 2 N }  \left( {1\over 2\epsilon_c - \omega - i\eta }+ {1\over 2\epsilon_c + \omega + i\eta } \right),
\end{multline}
where $\epsilon_c \propto {\cal N}_s^{-1} $ is the energy of the single-spinon state where the spinons condense
in the thermodynamic limit  ${\cal N}_s \to \infty$ and
$\chi_{SP;d}^{\mu \nu}({\bm q},\omega)$ is given in Eq.~\eqref{chispstat}.
We note that the polarization operator $\Pi({\bm q},\omega)$ contains a singularity at ${\bm q}={\bm 0}$ and $\omega=2\epsilon_c$  given by Eq.~(\ref{eq:pid}), namely
\begin{align}
\Pi_{\alpha_1 \alpha_2}&({\bm 0}, \omega) = f_{\alpha_1}^{{\bm Q}}(n,\bm 0) f_{\alpha_2}^{{\bm Q}}(n,\bm 0)  {{\cal N}_s \over 2 N } \nonumber \\ 
\times & \left( {1\over 2\epsilon_c - \omega - i\eta }+ {1\over 2\epsilon_c + \omega + i\eta } \right) +  \text{sub-leading terms}.
\end{align}
The  inverse of the matrix $\Pi_0 - \Pi({\bm 0}, 0)$, namely the propagator $D({\bm 0},0)$, 
satisfies 
\begin{equation}
\sum_{\alpha_1\alpha_2} D_{\alpha_1 \alpha_2}({\bm 0},2\epsilon_c) f_{\alpha_1}^{{\bm Q}} f_{\alpha_2}^{{\bm Q}} 
= {2 N  i \eta \over {\cal N}_s}
\end{equation} 
implying that
\begin{equation}
\chi_{FL;(a6)}^{\mu \nu}({\bm q},\omega) =  - \chi_{SP;d}^{\mu \nu}({\bm q},\omega).
\end{equation}
In other words,  the (singular)   SP contribution to the static and uniform magnetic susceptibility, represented by the diagram shown in Fig.~\ref{fig:chisp}~(d), is also canceled by the  diagram shown in Fig.~\ref{fig:chi1overN}~(a6).

\subsection{Self-energy and vertex corrections}

\begin{figure}[!t]
\centering
\includegraphics[scale=0.25]{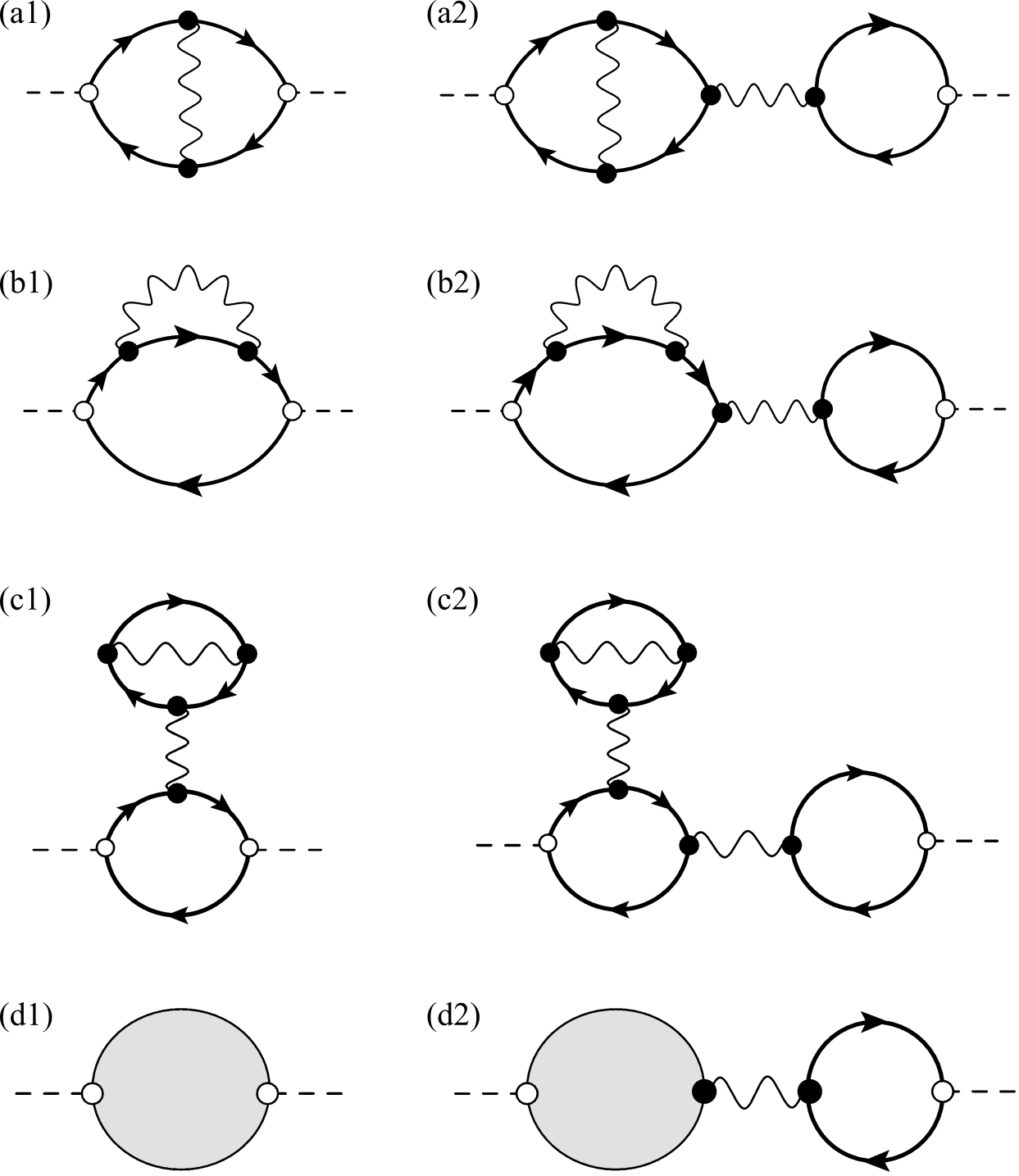}
\caption{Remaining  diagrams of nominal order $1/N$ (a1,b1,c1) and their corresponding counter diagrams (a2,b2,c2). (d1)  Arbitrary diagram of the $1/N$-expansion and  its counter diagram (d2).
The thick solid line represents the full single-spinon propagator including contributions from condensed and non-condensed spinons.}
\label{fig:counter_1overN}
\end{figure}

The three remaining  diagrams of nominal order $1/N$ that contribute to the dynamical spin susceptibility  are shown in Figs.~\ref{fig:counter_1overN} (a1), (b1) and (c1). 
The first diagram, ($a_1$), renormalizes the external vertex, while the other two, $(b1)$ and $(c1)$,
renormalize the single-spinon propagator. Like the saddle point contribution to the magnetic susceptibility, shown in Fig.~\ref{fig:chisp}), and 
the $1/N$ contribution shown in Fig.~\ref{fig:chi1overN}, these diagrams have poles 
at the {\it bare} single-spinon frequencies that must be canceled by higher order counter diagrams. 
By following exactly the same procedure that we used to demonstrate that the $1/N$ contribution shown in Fig.~\ref{fig:chi1overN} is a counter diagram for the saddle point contribution shown in Fig.~\ref{fig:chisp}, we can demonstrate that the counter diagrams  shown in  Figs.~\ref{fig:counter_1overN} (a2), (b2) and (c2) cancel the single-spinon poles of the contributions (a1), (b1) and (c1).
In general, the counter diagram of any diagram that has poles at the  bare single-spinon frequencies is constructed by simply adding 
the ``tail'' shown in Fig.~\ref{fig:counter_1overN}~(d).  It is interesting to note that  exactly the same procedure must be followed (even in absence of the condensate) to construct the counter diagram that cancels the SB density fluctuations~\cite{book2012interacting}. In other words, the same counter diagram simultaneously eliminates two unphysical effects: the residues of the single-spinon poles and density fluctuations that violate the constraint Eq.~\eqref{constraint}. As we discuss below, this qualitative improvement in the dynamical  spin susceptibility also leads to a significant quantitative improvement, which manifests in different aspects of the theory, such as the sum rule or the sensitivity of the results to the choice of $\alpha$ that are discussed in the next section.

\begin{figure}[!t]
\centering
\includegraphics[scale=0.1]{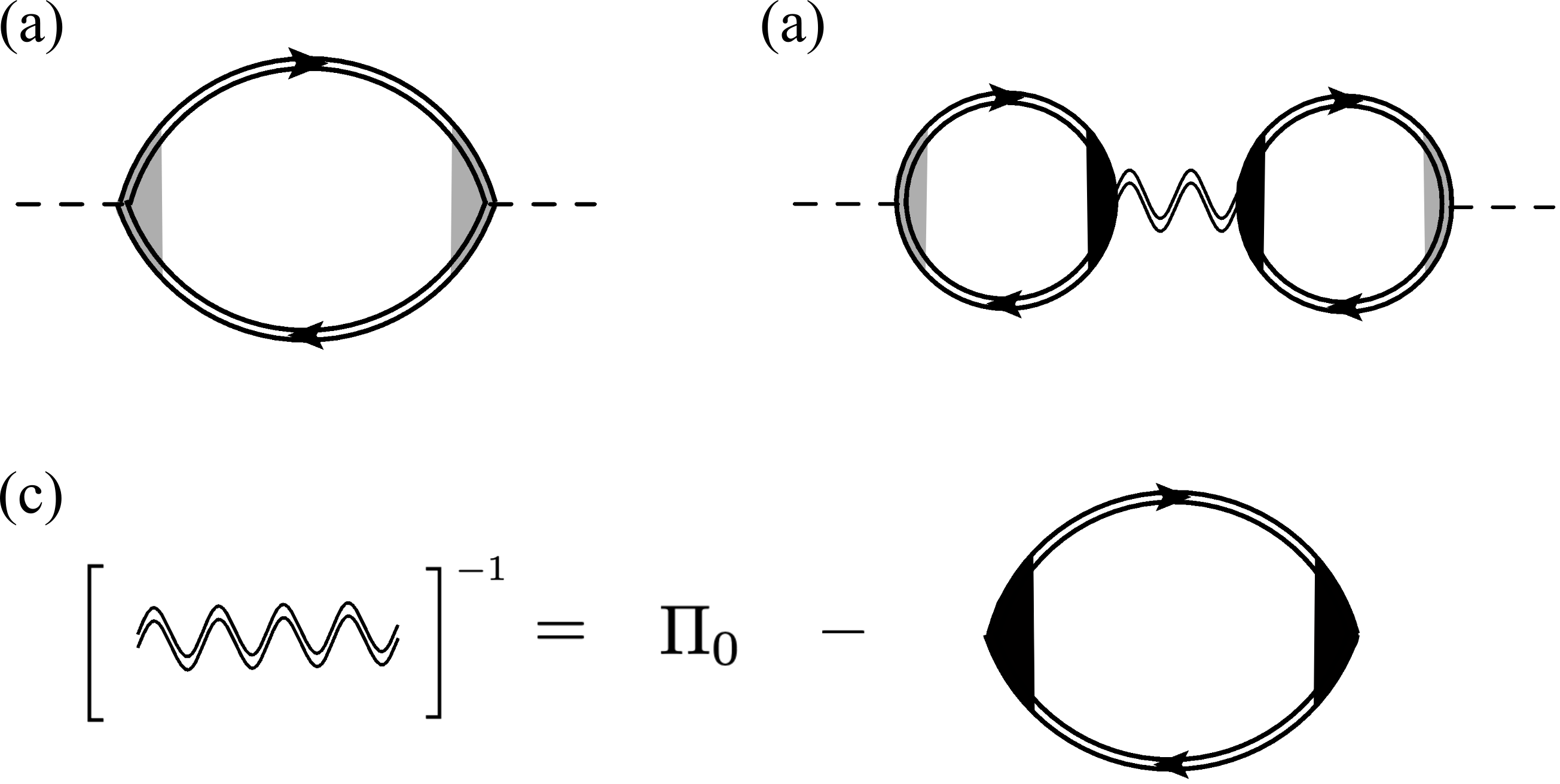}
\caption{(a) Magnetic susceptibility including 
renormalized external/internal vertices (gray/black shaded areas), spinon propagator (double solid line) and RPA propagator (double wavy line).
(b) Counter diagram that cancels out the single-spinon poles of the diagram shown in panel
 (a). (c) Renormalized propagator of the fluctuation fields.}
\label{fig:counter}
\end{figure}

For more general diagrams contributing to the magnetic susceptibility that include renormalized  external/internal vertex functions and propagators of the spinon and the fluctuation fields,
we can also construct counter diagrams that cancel the {\it renormalized} single-spinon poles. 
The renormalized magnetic susceptibility is  generally represented by the diagram shown in Fig.~\ref{fig:counter} (a). 
Once again, we can extend the previous derivation to demonstrate that the counter diagram of the renormalized bubble  is obtained by 
adding the renormalized tail shown in Fig.~\ref{fig:counter}~(b) and (c).
The demonstration is formally the same as the one given above in absence of vertex or self-energy renormalizations. The basic difference is the inclusion of the renormalization to the single-spinon Green's function for both non-condensed and condensed spinons, which leads to different quasi-particle energy and residues in comparison with the non-interacting case [see  Eq.~(\ref{green_n}) and Eq.~(\ref{green_c})]. By extending the previous discussion of the SP bubble diagram, we obtain a pole in the renormalized bubble diagram (Fig.~\ref{fig:counter} (a)), whose residue is formally given by Eq.~(\ref{residueSP}) after incorporating the single-spinon renormalizations.
Note that the demonstration of the cancellation is independent of  the  specific form of the external and the internal vertex functions. The crucial point is that the propagator of the fluctuation field must also be  renormalized, as shown in Fig.~\ref{fig:counter}~(c).

\begin{figure}[!t]
\centering
\includegraphics[scale=0.38]{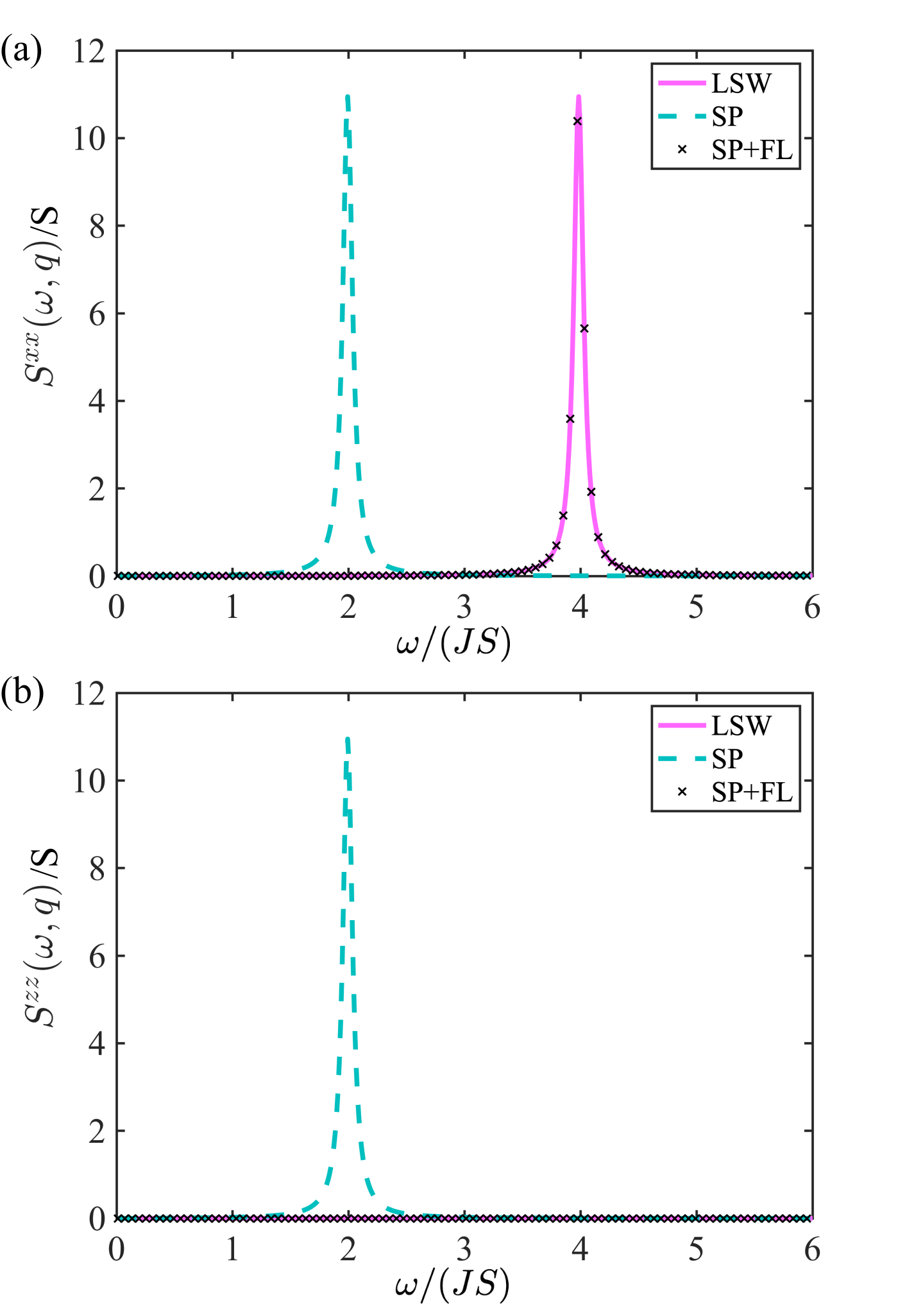}
\caption{Recovery of semiclassical limit ($S\rightarrow \infty$) with the Schwinger boson theory by including fluctuations around the SP solution $|G_3\rangle$ for a decoupling parameter $\alpha=0.5$.
Transverse (a) and longitudinal (b) DSSF for an arbitrarily chosen momentum ${\bm q}=(1.25, 2.09)$ in r.l.u. units.}
\label{fig:semiclassical}
\end{figure}

\section{Magnon poles}
\label{MP}

Along with the cancellation of the single-spinon poles, new poles emerge from the $1/N$ correction to the dynamical magnetic susceptibility.
The new poles arise from the tree-level diagrams of the RPA propagator $D({\bm q},\omega)$   shown in Fig.~\ref{fig:chi1overN} 
and they represent the magnons (true collective modes of the magnetically ordered ground state) as two-spinon bound states.
Unlike the single-spinon dispersion, which is strongly dependent on the mean-field decoupling of the original spin Hamiltonian (i.e. on the value of $\alpha$), we will demonstrate in this section that the single-magnon dispersion is much less sensitive to the mean-field decoupling.
The second purpose of this section is to demonstrate that the addition of the counter diagram shown in Fig.~\ref{fig:chi1overN} allows us to recover the linear SWT result in the $S \to \infty$ limit.


The RPA propagator of the square lattice Heisenberg antiferromagnet takes a block diagonal form $D_{A\lambda}({\bm q},\omega) \oplus D_B({\bm q},\omega)$ that can be explained by applying a symmetry argument to the expression of the polarization operator in terms of a retarded correlation function:
\begin{equation} 
4i \Pi_{\alpha_1\alpha_2}({\bm q},\omega)  = {1\over \sqrt{{\cal N}_s}} \sum_{{\bm r}} e^{-i {\bm q}\cdot {\bm r}}\int_{-\infty}^0 dt e^{i\omega t} 
 \langle G_{ i} \rvert [  {\hat \Phi}^{\alpha_1}_{{\bm r},0}, {\hat \Phi}^{\alpha_2}_{{\bm 0},t} ] \rvert G_{ i} \rangle,
\end{equation}
where  $i=1,2,3$ 
and $\alpha$ refers to the components of the vector field
\begin{align}
\hat{\boldsymbol \Phi}_{{\bm r},t} \equiv & \left((\alpha-1){\hat A}_{{\bm r}, {\bm \delta}}(t), (\alpha-1) {\hat A}_{{\bm r}, {\bm \delta}}^{\dagger} (t), \alpha {\hat B}_{{\bm r}, {\bm \delta}} (t),  \right. \nonumber \\
&  \left. -\alpha {\hat B}_{{\bm r}, {\bm \delta}}^{\dagger} (t), i {\hat n}_{\bm r}(t) \right), 
\end{align}
%
and  $\hat n_{\bm r} \equiv \hat {\bm b}_{\bm r}^{\dagger} \hat {\bm b}_{{\bm r}}^{}$ is the spinon density operator.
The {\it simple} BEC ground state  of the  mean-field Hamiltonian, $| G_3 \rangle$, is invariant under the product of
a global spin rotation by an angle $\varphi$ about the $z$-axis, ${\cal U}_z (\varphi)$, and a staggered gauge transformation $C(-\varphi)$,  $\tilde {\cal{U}}_z(\varphi)|G_3\rangle=|G_3\rangle$.
The gauge transformation is necessary to keep the \textit{simple} BEC state invariant because ${\cal U}_z (\varphi)$ maps the macroscopically occupied single-particle state with ${\bm q}={\bm 0}$  into a condensate in a single-particle mode that is a linear combination of the  ${\bm q}={\bm 0}$ and ${\bm \pi}$ modes [see Eq.\eqref{eq:u1c_q}].
The operators 
${\hat A}_{{\bm r},{\bm \delta}}^{}$, 
${\hat A}_{{\bm r},{\bm \delta}}^{\dagger}$, and ${\hat n}_{\bm r} $ remain invariant
under the gauge transformation  $C(\varphi)$, while the operators 
${\hat B}_{{\bm r},{\bm \delta}}^{}$ and  
${\hat B}_{{\bm r}, {\bm \delta}}^{\dagger}$, 
acquire a complex phase factor,
\begin{equation}
    \hat B_{\bm r,\bm \delta}^{\dag} \ \overset{C(\varphi)}{\longrightarrow} \  e^{i\eta_{\bm r} \varphi} \ \hat B_{\bm r,\bm \delta}^{\dag},
    \label{eq:gaugeB}
\end{equation}
implying that, for instance
\begin{align*}
    \langle G_{i} \rvert {\hat A}^{\dag}_{\bm 0, \bm \delta} {\hat B}^{\dag}_{\bm r, \bm \delta} \rvert G_{i} \rangle
    = & \ \langle G_{i} \rvert \tilde {\cal{U}}_z^\dag(\varphi) {\hat A}^{\dag}_{\bm 0, \bm \delta} {\hat B}^{\dag}_{\bm r, \bm \delta} \tilde {\cal{U}}_z(\varphi) \rvert G_{i} \rangle \\
    = & \ e^{i\eta_{\bm r} \varphi} \langle G_{i} \ \rvert  {\hat A}^{\dag}_{\bm 0, \bm \delta} {\hat B}^{\dag}_{\bm r, \bm \delta}  \rvert G_{i} \rangle.
\end{align*}
Then the cross correlation function between the ${\hat A}$ or ${\hat n}$ and ${\hat B}$ operators must be zero and the  propagator matrix has the form $D({\bm q},\omega)=D_{A\lambda}({\bm q},\omega) \oplus D_B({\bm q},\omega)$. 

The propagator $D_{A\lambda}$ includes a zero-energy mode for each $\omega$ and ${\bm q}$ value  because of the redundant gauge degree of freedom (the zero mode corresponds to a gauge transformation $b_{i\sigma}(\tau) \rightarrow b_{i\sigma}(\tau) e^{i \theta_{i}(\tau)}$). This unphysical mode does not contribute to any physical observable.
$D_{A\lambda}$ does not have isolated poles.
By contrast, the propagator of the $B$-fields, $D_{B}$, includes a pole singularity that becomes gapless at  
${\bm q}={\bm 0}$ and ${\bm \pi}$. As we will see below, these poles represent  the  two Goldstone modes (transverse magnons) at 
${\bm q}={\bm 0}$ and ${\bm \pi}$ expected for \emph{collinear} magnetic ordering (it breaks two continuous {spin rotation} symmetries).


{\color{red}
\begin{figure*}[!t]
\centering
\includegraphics[scale=0.59]{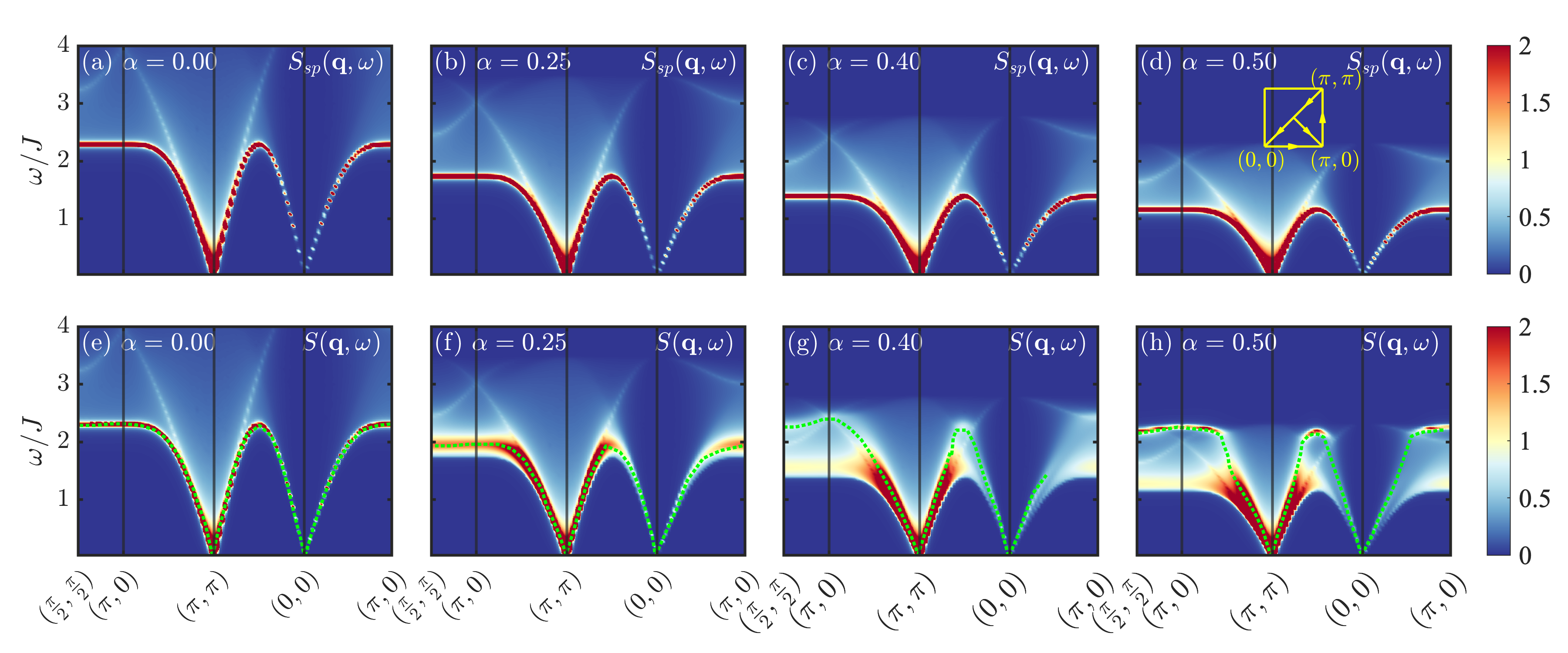}
\caption{Transverse dynamical spin structure factor ($S=1/2$). (a-d) show the SP result for the simple BEC solution represented by the state  $|G_3\rangle$  and $\alpha = 0, 0.25, 0.4, 0.5$, while (e-h) include the $1/N$ correction  described in Fig.~\ref{fig:chi1overN}. Note that $S^{xx}({\bm q},\omega)=S^{yy}({\bm q},\omega)$. The green dots indicate the magnon dispersion. Inset: high-symmetry path of horizontal axis.}
\label{fig:trans_sqw}
\end{figure*}

\begin{figure*}[!t]
\centering
\includegraphics[scale=0.59]{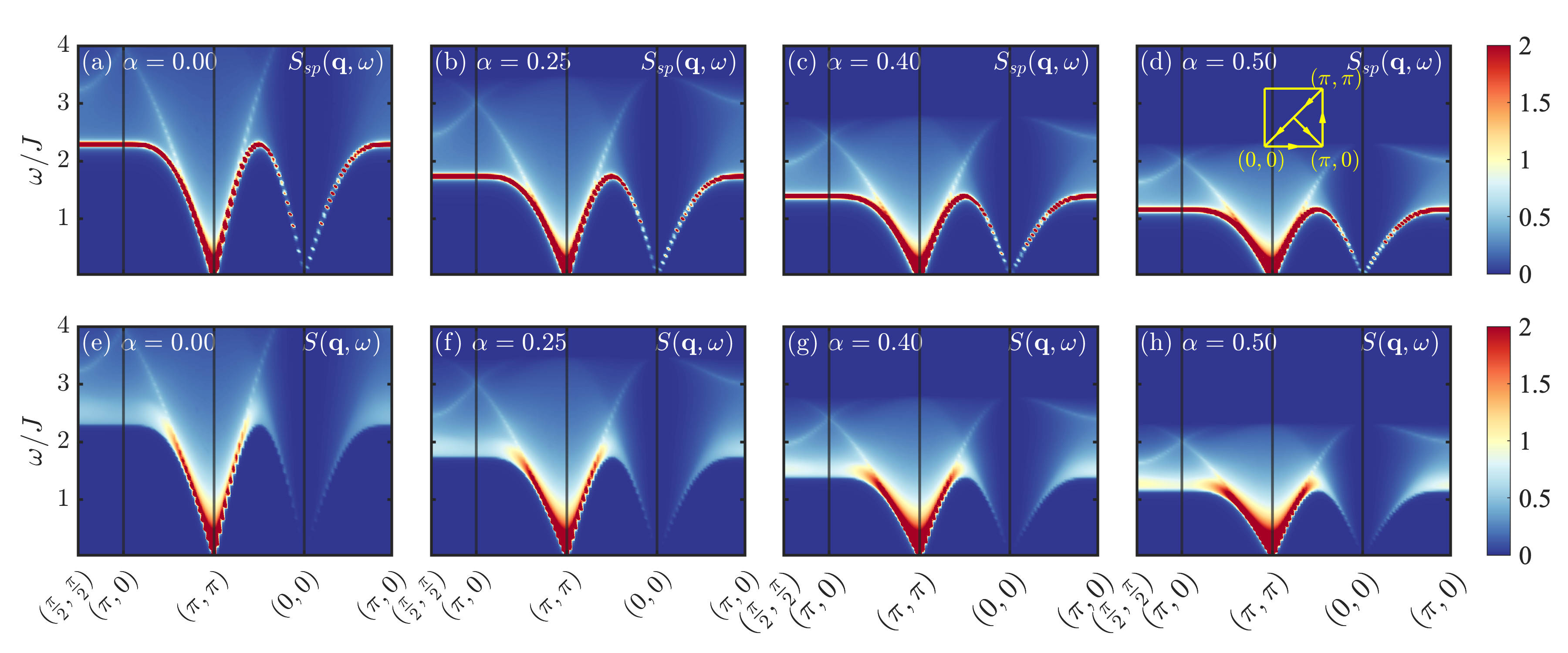}
\caption{Longitudinal dynamical spin structure factor  ($S=1/2$). (a-d) show the SP result for the simple BEC solution represented by the state  $|G_3\rangle$ and $\alpha = 0, 0.25, 0.4, 0.5$, while (e-h) include the $1/N$ correction described in Fig.~\ref{fig:chi1overN}.  Inset: high-symmetry path of horizontal axis.}
\label{fig:long_sqw}
\end{figure*}

\begin{figure*}[!t]
\centering
\includegraphics[scale=0.59]{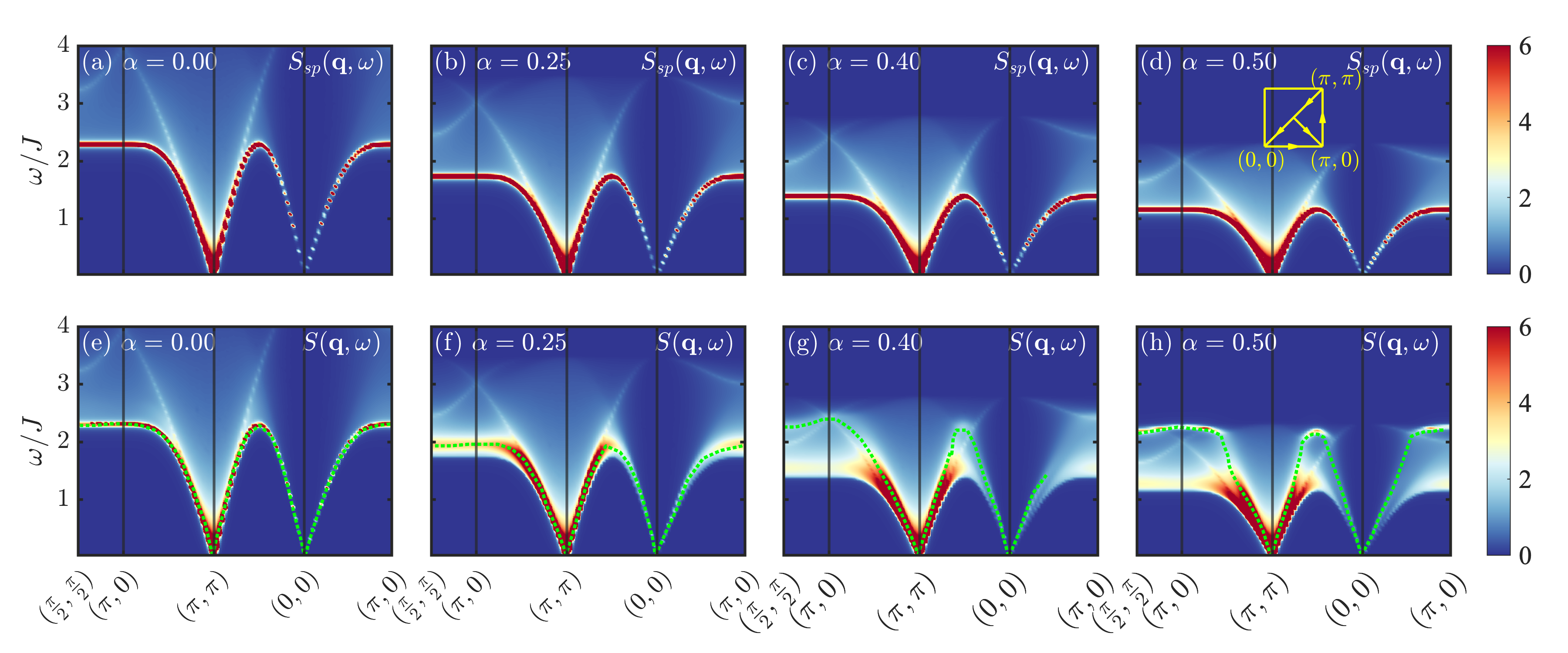}
\caption{Total dynamical spin structure factor  $S({\bm q},\omega) = \sum_{\mu} S^{\mu \mu} ({\bm q},\omega)$ ($S=1/2$). (a-d) show the result for the SP solution $|G_3\rangle$ and $\alpha = 0, 0.25, 0.4, 0.5$, and (e-h) include the $1/N$ correction described in Fig.~\ref{fig:chi1overN}. The green dots indicate the magnon dispersion. Inset: high-symmetry path of horizontal axis.}
\label{fig:trace_sqw}
\end{figure*}
}

Let us consider now the dynamical magnetic susceptibility with the correction depicted in Fig.~\ref{fig:chi1overN}. 
As we explained in the previous section, we only include the diagrams shown in Fig.~\ref{fig:chi1overN} (a1-a6) because they are 
the counter diagram of the SP contribution shown in Fig.~\ref{fig:chisp}.
Explicitly, the $1/N$ correction shown in Fig.~\ref{fig:chi1overN} is given by
\begin{equation}
\chi_{FL}^{\mu \nu} ({\bm q}, \omega) = \sum_{\alpha_1 \alpha_2} \Lambda^{\mu \alpha_1} ({\bm q},\omega) {1 \over N} D_{\alpha_1 \alpha_2} ({\bm q},\omega)  \Lambda^{\nu \alpha_2} (-{\bm q}, -\omega) ,
\end{equation}
where 
\begin{equation} 
2i \Lambda^{\mu \alpha} ({\bm q},\omega) = {1\over \sqrt{{\cal N}_s}} \sum_{{\bm r}} e^{-i {\bm q}\cdot {\bm r}}\int_{-\infty}^0 dt e^{i\omega t} 
 \langle G_{ i} \rvert [  {\hat S}^{\mu}_{{\bm r},0}, {\hat \Phi}^{\alpha}_{{\bm 0},t} ] \rvert G_{ i} \rangle,
\end{equation}
is the cross susceptibility between the spin operator ${\hat S}_{\bm r}^{\mu}$ and the fluctuation field component $\hat \Phi_{{\bm r},t}^{\alpha}$ at the  SP level.

Since $|G_1\rangle$ and the bond operators are invariant under global spin  SU(2) spin rotations, while the spin operators transform like vectors, the cross susceptibility $\Lambda^{\mu \alpha} ({\bm q},\omega)$ is equal to zero, implying that $\chi_{FL}^{\mu \nu} ({\bm q}, \omega)=0$ for the isotropic SP solution $|G_1\rangle$. 

The SP solution $|G_2\rangle$ is only invariant under the residual U(1) symmetry  subgroup of global rotations ${{\cal U}}_{\bm z} (\varphi)$ along the $z$-axis (direction of the ordered moments). The transverse spin components ${\hat S}_{\bm r}^{\pm} = {\hat S}_{\bm r}^{x} \pm i {\hat S}_{\bm r}^{y} $ acquire a complex phase $\exp(\pm i \varphi)$ under these  rotations. Consequently, we still obtain that $\langle G_{2} \rvert {\hat S}^{\pm}_{\bm r} {\hat \Phi}^{\alpha}_{\bm 0} \rvert G_{2} \rangle = 0$, implying that $\Lambda^{\mu \alpha} ({\bm q},\omega)=0$ and $\chi_{FL}^{\mu \nu} ({\bm q}, \omega)=0$ for the transverse components $\mu, \nu=x,y$. 
 In addition, the invariance of
$\rvert G_{2} \rangle$ under the gauge transformation $C(\varphi)$ implies that  $\langle G_{2} \rvert {\hat S}^{z}_{\bm r} {\hat B}^{\dag}_{\bm 0, \bm \delta} \rvert G_{2} \rangle = 0$.  On the one hand, since the poles of the  RPA propagator arise from the $B$-sector ($D_B$),  the longitudinal susceptibility does not acquire new poles after adding the correction due to fluctuations. On the other hand, the 
finite correction  due to the non-zero cross correlation function between  ${\hat S}^{z}_{\bm r}$ and the $A-\lambda$ fields leads to the cancellation of the single-spinon poles of the SP contribution
$\chi_{SP}^{\mu \nu} ({\bm q}, \omega)$.
This is the qualitatively correct result for the longitudinal component of the spin susceptibility for the collinear  antiferromagnet under consideration (the magnons are  transverse modes).

The situation is qualitatively different for the simple condensate $|G_3\rangle$ as it is reveled by the  following symmetry argument. 
$|G_3\rangle$ is invariant under the residual U(1) symmetry group of global spin rotations about the $z$-axis \emph{up to a gauge transformation}:  $ {\tilde {\cal U}}_{\bm z} = {\cal U}_{\bm z} (\varphi)  C(-\varphi) \rvert G_{3} \rangle = \rvert G_{3} \rangle$, implying that  $\langle G_{3} \rvert {\hat S}^{z}_{\bm r} {\hat B}^{\dag}_{\bm 0, \bm \delta} \rvert G_{3} \rangle = 0$ because ${\hat S}^{z}_{\bm r}$ is invariant under that transformation, while ${\hat B}^{\dag}_{\bm 0, \bm \delta}$ acquires a phase factor [see Eq.~\eqref{eq:gaugeB}]. In addition,
$\langle G_{3} \rvert {\hat S}^{\pm}_{\bm r} {\hat \Phi}^{\alpha}_{\bm 0} \rvert G_{3} \rangle = 0$ for ${\hat \Phi}^{\alpha}= {\hat A}, \hat {A^{\dag}}, {\hat n}$ because these operators are invariant under ${\tilde {\cal U}}_{\bm z}$, while the transverse spin components ${\hat S}^{\pm}_{\bm r}$ acquire a phase. 
Importantly, for the transverse channel, the phase factor acquired by $\hat S_{\bm r}^{\pm}$ due to the global spin rotation ${\cal U}_{\bm z} (\varphi)$
is compensated by the phase factor acquired by $\hat B_{\bm r,\bm \delta}^\dag$ due to the gauge transformation $C(-\varphi)$, allowing for  a non-zero value of $\Lambda^{\mu \alpha}$. \emph{Consequently, the simple BEC SP solution $|G_3\rangle$ is the only one that has a finite correction due to fluctuations (diagram of Fig.~\ref{fig:chi1overN}) to the transverse components of the susceptibility:  $\chi_{FL}^{\mu \nu} ({\bm q}, \omega) \neq 0$  for $\mu, \nu=x,y$}. 
In summary, $\Lambda^{\mu \alpha}$ remains finite \emph{both for the longitudinal and for the transverse components}.
In particular, the longitudinal $z$ component of the spin has a finite cross correlation function with the 
$A-\lambda$ fields, while the transverse spin components  have a finite cross correlation function with the $B$ fields.
The pole of the RPA propagator  $D_B$ gives rise to a $\delta$-peak in the transverse spin channel of the DSSF, while the correction due to fluctuations in the longitudinal channel cancels out the poles of $\chi_{SP}^{z z} ({\bm q}, \omega) \neq 0$ and leaves a two-spinon continuum associated with the branch-cut singularity of $D_{A\lambda}$, which is the qualitatively correct result. 
 
This qualitative distinction between $\rvert G_2 \rangle$ and $\rvert G_3 \rangle$ holds true for any collinear antiferromagnet revealing that the simple condensate $\rvert G_3 \rangle$ provides a more adequate  description of the magnetically ordered state because a finite contribution from the fluctuations in the transverse channel is essential to cancel out the single-spinon poles of the SP contribution for $\alpha \neq 0$.

The spectral weight of the two-spinon continuum vanishes in the $S\rightarrow \infty$ limit   
and the spectral weight of the  DSSF  is restricted to the $\delta$-peak associated with the single-pole 
in the transverse spin channel (magnon modes).
As for the triangular lattice case~\cite{zhang2019large}, the dispersion and the spectral weight of these magnons become identical to the linear SWT result in that limit (see Fig.~\ref{fig:semiclassical}).
Moreover, the recovery of the linear SWT result in the large-$S$ limit is independent to the mean-field decoupling of the original spin Hamiltonian, i.e., of the choice of $\alpha$ in Eq.~\eqref{eq:SBlargeN}.

Fig.~\ref{fig:trans_sqw} shows  the \emph{transverse} DSSF along  high-symmetry directions for different values of $\alpha$, i.e., different decoupling schemes that lead to different simple BEC SP solutions. The upper panels correspond to the SP result, while the lower panels include the contribution from the counter-diagram depicted in Fig.~\ref{fig:chi1overN}.
As we explained above, the contribution from the counter-diagram vanishes for $\alpha=0$ because of the lack of $B$-fields in that particular decoupling scheme. In this particular and fortuitous case, the single-spinon dispersion coincides with the single-magnon dispersion: $\varepsilon_{\bm{q}} = \omega_{\bm q}$.
The situation is very different for finite values of $\alpha$, where the presence of $B$-fields leads to the above-mentioned cancellation of the single-spinon poles of the SP solution and to the emergence of new magnon poles arising from the fluctuation of the $B$-fields. It is interesting to note that the bandwidth of the single-spinon dispersion (poles of the SP solution) decreases by a factor of two when $\alpha$ evolves from zero to $0.5$ (see Fig.~\ref{fig:trans_sqw}(a)-(c)). In contrast, the bandwidth of the single magnon dispersion is much less dependent on $\alpha$ (see Fig.~\ref{fig:trans_sqw}(d)-(f)). For instance the energy of the 
 magnon at ${\bm q}=(0,\pi)$
varies over the range  $[1.8, 2.316]$ for $0 \leq \alpha \leq 0.5$. This energy range is consistent with numerical results~\cite{powalski2015roton,powalski2018mutually}.
In other words, the inclusion of corrections beyond the SP make the final result less sensitive to the choice of the SP. Nevertheless, there is still a significant $\alpha$-dependence of the magnon spectral weight and the distribution of the two-spinon continuum, implying  that for low-order expansions in $1/N$ it is necessary to introduce a criterion for choosing an optimal decoupling scheme (value of $\alpha$). In particular, note that some magnons are overdamped above a critical value of $\alpha$ because they overlap with the two-spinon continuum
 [e.g., the magnons modes with energy $\sim 2J$ along the path $(\pi/2,\pi/2)$-$(\pi,0)$ shown in 
 Figs.~\ref{fig:trans_sqw} (e,f,g)].

For the longitudinal component of the DSSF, shown in Fig.~\ref{fig:long_sqw}, the main difference between the SP result (upper row)
and the one that includes the contribution from the counter-diagram shown in  Fig.~\ref{fig:chi1overN}
is that the latter one (lower row) has no poles. 
As we already explained, the lack of poles in the longitudinal channel is the qualitatively correct result. It is also important to note that the distribution of spectral weight in the two-spinon continuum is still strongly dependent on $\alpha$. This is a direct consequence of the $\alpha$-dependence of the single-spinon
dispersion at the SP level: $\varepsilon_{\bm{q}} = (1-\alpha) \omega_{\bm q}$.
In order to make this result less sensitive to $\alpha$, it is necessary to renormalize the single-spinon propagator by including diagrams such as the ones shown in Fig.~\ref{fig:counter_1overN}(b1) and (c1). This renormalization is also expected to account for many-body effects that are not captured at the current level of approximation. For instance, it is well-known that the hybridization between single and three-magnon states leads to  a roton-like anomaly in the single-magnon dispersion at ${\bm q}=(0,\pi)$ or ${\bm q}=(\pi,0)$~\cite{Sylju2000,Ronnow01,Piazza2015,powalski2015roton,powalski2018mutually}. Since magnons are obtained as two-spinon bound states in the SBT, few-body effects caused by three-magnon states will manifest via hybridization between the two-spinon and six spinon sectors (note that the renormalization of the single-spinon propagator will in turn renormalize the RPA propagator and its poles).



\begin{figure*}[!t]
\centering
\includegraphics[scale=0.32]{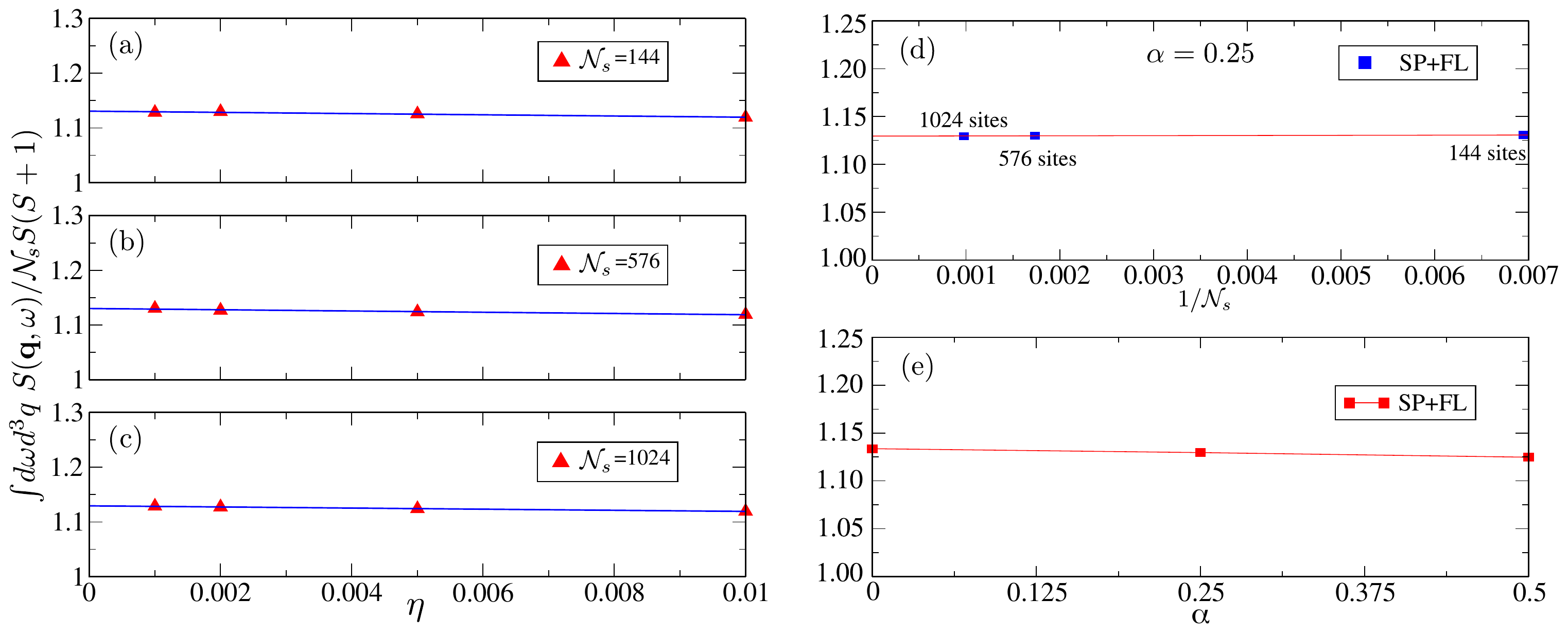}
\caption{ Linear extrapolation of the integral over frequency and momentum of the dynamical spin structure factor after correcting the simple BEC SP  result [see Eq.~\eqref{srsp}] with the contribution described in Fig.~\ref{fig:chi1overN}. The panels (a), (b) and (c) correspond to extrapolations as a function of the width $\eta$ for $\alpha=0.25$ and finite lattices of ${\cal N}_s=144$, ${\cal N}_s=576$ and ${\cal N}_s=1026$ respectively. Panel (d) shows the finite size scaling and the extrapolation to the thermodynamic limit of the extrapolated ($\eta \to 0$) results obtained in the previous panels. 
Panel (e) shows the final result after performing the $\eta \to 0$ and ${\cal N}_s \to \infty$ extrapolations for different values of $\alpha$.
}
\label{fig:sum_rule}
\end{figure*}

Fig.~\ref{fig:trace_sqw} shows the  trace of the DSSF (transverse plus longitudinal components) for different values of $\alpha$.  As we anticipated in Sec.~\ref{SP}, the exact  sum rule  $ \int d \omega d^3q S({\bm q},\omega) = {\cal N}_s S(S+1)$ provides a good test for any approximation scheme. According to Eq.~\eqref{srsp},  the SP result overestimates the sum rule by $\approx 50\%$. In contrast, the integration of the total dynamical spin structure factor, shown in Fig.~\ref{fig:sum_rule}, gives $\int d \omega d^3q S({\bm q},\omega) = {\cal N}_s \zeta(\alpha) S(S+1)$, with
$\zeta(0) \approx 1.133$ and $\zeta(0.5) \approx 1.125$. To obtain these results, we computed the integral of $S({\bm q},\omega)$ for different values of $\eta$ (Lorentzian broadening) and for system sizes ${\cal N}_s=144,\ 576$ and $1024$. We then performed a linear extrapolation  in the width $\eta \to 0$ (see Fig.~\ref{fig:sum_rule}~(a)-(c)) for each system size, and the result was in turn extrapolated to the thermodynamic limit $1/{\cal N}_s \to 0$.
As it is clear from Fig.~\ref{fig:sum_rule}, the sum rule exhibits a very weak dependence on $1/{\cal N}_s$ and $\eta$ for small enough values of these parameters. As it is shown in  Fig.~\ref{fig:sum_rule}~(e),
the value of $\zeta(\alpha)$ for intermediate values of $\alpha$, $0< \alpha < 0.5$, is well approximated by a linear interpolation between the above-mentioned extreme values. Clearly, the sum rule remains basically independent of the decoupling scheme and the
inclusion of the new contribution produces a much better approximation to the exact sum rule.
Since violations of the sum rule arise from violations of the physical constraint Eq.~\eqref{constraint}, we confirm that the addition of the counter-diagram shown in Fig.~\ref{fig:chi1overN} also leads to a much better approximation of the local constraint.

Finally, we should say a few words about the criterion that must be adopted to determine the ``optimal'' decoupling scheme, i.e., the value of $\alpha$. In the past, different groups have used comparisons of the ground state energy against exact diagonalization results on finite size clusters to discriminate between different decoupling schemes~\cite{auerbach1988spin,Yoshioka1991,Ceccatto1993,Gazza1993,Ceccatto1994,Trumper1997,Manuel1999}. An alternative criterion could be to directly compare the DSSF against results obtained on finite size clusters or from experiments. 
This criterion has been applied to the  triangular lattice antiferromagnet obtaining an optimal value of $\alpha$ close to $0.5$~\cite{ghioldi2018dynamical,zhang2019large}.

\section{Summary}
\label{SUM}

As we already mentioned in the introduction, most attempts of using the SBT for describing magnetically ordered systems  have not gone beyond the saddle-point level.
In this work we have presented a comprehensive approach to the selection of the magnetically ordered SP solution and inclusion of higher order corrections of the $1/N$ expansion.
As we argued in the previous sections, the presence of a condensate associated with the magnetic ordering generates multiple subtleties that have been omitted in previous discussions of the SBT. The main motivation of this work has not been to solve again the particular case of the square lattice antiferromagnet, but to employ this textbook  model to illustrate how a proper choice of the SP solution and a careful analysis of the cancellations that occur in the presence of a condensate can improve the results both qualitatively and quantitatively.

On the one hand, we have carefully examined the consequences of the freedom associated with the choice of the decoupling scheme (value of $\alpha$) and the fragmented vs. simple nature of the BEC ground state 
of the SP Hamiltonian for a fixed value of $\alpha$.
On the other hand, we have seen that Feynman diagrams, which in absence of the condensate would contribute to a given order in $1/N$, acquire singular lower order contributions  in the presence of a condensate. Upon expanding the dynamical spin susceptibility in powers of $1/N$, these singular contributions lead to the existence of counter-diagrams that exactly cancel out the residues of the single-spinon poles of a given diagram, as expected for the excitation spectrum of magnetically ordered states. The true quasi-particles of the ordered system are then magnons or transverse modes that to the lowest non-trivial order in $1/N$ arise from poles of the RPA propagator. Correspondingly, a main conclusion of this work is that in a  correct  expansion of the magnetic susceptibility of an ordered magnet \emph{each diagram must  be accompanied by its counter-diagram}. 

Among other consequences, this important conclusion implies that it is strictly necessary to go  beyond the SP level  to obtain physically correct results for  magnetically ordered systems. 
In other words, the loop diagram depicted in Fig.~\ref{fig:chisp}, which is the SP contribution to the dynamical spin susceptibility, must be accompanied by its counter diagram depicted in Fig.~\ref{fig:counter_1overN}, which is \emph{nominally} of order $1/N$. This requirement was first noticed (without a clear justification) in a previous work on the triangular Heisenberg antiferromagnet~\cite{ghioldi2018dynamical}. The reason why this subtle point remained hidden for approximately 20 years is that the SBT was first applied to the square lattice Heisenberg antiferromagnet~\cite{auerbach1988spin,Arovas1988}. As we discussed in the previous section, the $AA$ or $\alpha=0$ decoupling scheme  that was originally applied to the square lattice Heisenberg antiferromagnet~\cite{auerbach1988spin,Arovas1988} has an important peculiarity: the single-spinon dispersion coincides with the single-magnon dispersion and the contribution from the counter-diagram vanishes \emph{for the transverse} dynamical spin susceptibility.
This ``pathology'' of the particular case $\alpha=0$ led the community to believe that the SP level of the SBT is enough to obtain the true collective modes of magnetically ordered states.
While this generalized belief was questioned after noticing the impossibility of recovering the correct large-$S$ limit with the SP result
for the general case (non-collinear orderings)
~\cite{Chandra1991,Coleman1994}, the origin of this failure was not properly understood. As we have demonstrated here, the identity between the single-spinon and single-magnon dispersions is not true for general decoupling schemes ($\alpha\neq 0$).
Moreover, the pathology is completely absent in systems with non-collinear magnetic ordering, such as the triangular lattice AFM.
As it was demonstrated in a previous work~\cite{zhang2019large}, the addition of the counter diagram allows us to recover the correct result in the large-$S$ limit. Furthermore, as it was shown in this work, \emph{this result is independent of the decoupling scheme parametrized by $\alpha$}.

Another subtlety of the SBT applied to the square lattice Heisenberg antiferromagnet is related to the freedom associated with the choice of the spinon condensate in the presence of a symmetry breaking field. This freedom arises from the collinear nature of the antiferromagnetic ordering that leads to a U(1) residual symmetry group for SU(2) invariant Heisenberg interactions.
While the symmetry breaking field polarizes the condensed spinons along the field direction in the twisted reference frame, the spinons can still  condense in more than one gapless mode (${\bm q}={\bm 0}$ or ${\bm q}= {\bm\pi}$ for a particular gauge choice of the square lattice antiferromagnet SBT considered here). This  remaining ground state degeneracy of the SP Hamiltonian introduces an extra freedom in the choice of the condensate. 
As we have seen in Sec.~\ref{SP}, the 
 importance of choosing a ``simple'' spinon BEC  instead of the fragmented BEC  adopted in previous works~\cite{book2012interacting} becomes evident after including corrections beyond the SP level. Furthermore, if we continuously deform the  triangular lattice antiferromagnet  into the square lattice Heisenberg antiferromagnet, the \emph{unique} BEC solution of the non-collinear triangular antiferromagnet evolves continuously into the simple BEC solution of the square lattice antiferromagnet. 

By considering all the above-mentioned subtleties, we have finally shown that including corrections beyond the SP level is crucial not only to reveal the true quasi-particles of the magnetically ordered state, but also to reduce the dependence of the  dynamical spin susceptibility on the decoupling scheme (value of $\alpha$).
We note however that the two-spinon continuum is still strongly dependent on $\alpha$ because we have not included the diagrams shown in Fig.~\ref{fig:counter_1overN}(b1) and (c1) that renormalize the single-spinon propagator. A natural consequence of this limitation of the current level of approximation is that some magnon modes become overdamped  above a critical value of $\alpha$ because they overlap with the two-spinon continuum. We expect that this undesirable result will be corrected after including a proper renormalization of the single-spinon propagator.
A detailed study of the different effects of the single-spinon renormalization will be the subject of future work. 
Based on the preliminary results presented in this work, we conclude that the optimal decoupling scheme (value of $\alpha$) should be determined by comparing the spectrum of low-energy excitations against some reference that can either be a numerical result on finite lattices or an experimental result. 

Finally, the SBT presented here is particularly relevant for modelling magnetically ordered materials in the proximity of a quantum critical point that signals a continuous transition in a spin liquid state. The simple reason is that magnons become composite particles (two-spinon bound states) in the proximity of the ``quantum melting point''. The triangular lattice Heisenberg antiferromagnet~\cite{ghioldi2018dynamical,Scheie21} provides a natural example of this scenario because the quantum melting point is reached by adding a second neighbor antiferromagnetic interaction that is only 6\% of the nearest neighbor exchange~\cite{Zhu_2018,Maksimov_2019,Hu15,Iqbal16,Saadatmand16,Wietek17,Gong17}.

\begin{acknowledgments}
We wish to thank A. V. Chubukov, C. Gazza and O. Starykh for useful conversations. This work was partially supported by CONICET under grant Nro 364 (PIP2015).
E.~A.~G. was partially supported from the LANL Directed Research and Development program. The work of C.D.B was supported by  the U.S. Department of Energy, Office of Science, Office of Basic Energy Sciences, under Award No.~DE-SC0022311.
\end{acknowledgments}


\appendix

\section{Relation from particle-hole symmetry}

The Nambu representation has an artificial ``particle-hole'' symmetry because $\psi_{i}=P\left(\psi_{i}^{\dagger}\right)^{T}$
or $\psi_{\bm q}=P\left(\psi_{-{\bm q}}^{\dagger}\right)^{T}$ (momentum space), where $P=\sigma_{x}\otimes I_{0}$. 
The invariance of the effective action under this transformation implies that the dynamical matrix satisfies $P{\cal M}^T(-p, -k)P^T = {\cal M}(k,p)$,
where $k\equiv ({\bm k},i\omega)$ and $p\equiv ({\bm p},i\nu)$. 
The external/internal vertices are the first derivatives of ${\cal M}(k,p)$ with respect to the external source term $h^{\mu}_{q}$ or the fluctuation fields $\phi_{\alpha}(q)$. 
It then follows that
\begin{align}
P\left( u^{\mu} \right)^T P &= u^{\mu}, & P\left( v_{\alpha} ({\bm k},{\bm q}) \right)^T P &= v_{\alpha} (-{\bm q},-{\bm k}).\label{A1}
\end{align}
In the SP approximation, the inverse of ${\cal M}(k,p) \equiv (-i\omega \gamma^0 + H_{sp}({\bm k})) \delta_{k,p}$ gives rise to the single spinon's Green's function. We have
\begin{align}
P H_{sp}^*(-{\bm k}) P &= H_{sp}({\bm k}), & P {\cal G}_n(-{\bm k},-\omega)^T P = &{\cal G}_n({\bm k},\omega).
\label{A2}
\end{align}
The eigenstates of the SP Hamiltonian must then satisfy the relations
\begin{align}
X_{{\bm k},\alpha} &= P {\bar X}^*_{-{\bm k},\alpha}, & {\bar X}_{{\bm k},\alpha} &= P {X}^*_{-{\bm k},\alpha}.
\end{align}
In particular, the eigenstate $\rvert c_{{\bm Q}}\rangle \equiv X_{{\bm Q},+1} $ or ${\bar X}_{{\bm Q},+1}$ (identical in the thermodynamic limit) satisfies
\begin{align}
 \rvert c_{\bm Q} \rangle = P \left( \rvert c_{\bm Q} \rangle \right)^*,\label{A3}
\end{align}
where we have assumed that ${\bm Q} = -{\bm Q}$ is an inversion invariant momentum vector.

Using Eqs.~(\ref{A1})-(\ref{A3}), we can show that the two diagrams in Fig.~\ref{fig:chisp} (b) and (c) are equal to each other. For instance, their contributions to the
DSSF, Eqs.~(\ref{eq:chisp}) and (\ref{eq:chisp2}), are identical. Similar conclusion holds in the calculation of the other diagrams in this work.

\bibliographystyle{ieeetr}
\bibliography{ref}

\end{document}